\documentclass[10pt,journal,compsoc]{IEEEtran}

\ifCLASSOPTIONcompsoc
  \usepackage[nocompress]{cite}
\else
  \usepackage{cite}
\fi

\usepackage{tikz,xcolor,hyperref}
\DeclareRobustCommand{\orcidicon}{%
	\begin{tikzpicture}
	\draw[lime, fill=lime] (0,0) 
	circle [radius=0.16] 
	node[white] {{\fontfamily{qag}\selectfont \tiny ID}};
	\draw[white, fill=white] (-0.0625,0.095) 
	circle [radius=0.007];
	\end{tikzpicture}
	\hspace{-2mm}
}
\foreach \x in {A, ..., Z}{%
	\expandafter\xdef\csname orcid\x\endcsname{\noexpand\href{https://orcid.org/\csname orcidauthor\x\endcsname}{\noexpand\orcidicon}}
}

\usepackage{amsmath,amssymb,amsfonts, amsthm, wasysym, stmaryrd}
\usepackage{algorithmic}
\usepackage{graphicx}
\usepackage{textcomp}

\usepackage{multirow}
\usepackage{array}
\usepackage{longtable}
\usepackage{multicol}
\usepackage{pifont}
\usepackage[justification=centering]{caption}
\usepackage[T1]{fontenc}
\usepackage{enumerate}
\usepackage[shortlabels]{enumitem}

\usepackage{tcolorbox}

\usepackage{afterpage}
\usepackage{lscape}
\usepackage{arydshln}
\newif\ifblackandwhite
\blackandwhitetrue
\usepackage{etoolbox}
\usepackage{longtable}%
\usepackage{supertabular,booktabs}

\usepackage{xcolor,colortbl}
\usepackage{dcolumn}

\usepackage{pdflscape}
\usepackage{oplotsymbl}
\usepackage{rotating}

\ifblackandwhite
  
\else

\fi

\usepackage{tikz}
\usetikzlibrary{shapes.geometric}

\usepackage{xurl}

\makeatletter
\newcommand{\ccell}[3][]{%
  \kern-\fboxsep
  \if\relax\detokenize{#1}\relax
    \expandafter\@firstoftwo
  \else
    \expandafter\@secondoftwo
  \fi
  {\colorbox{#2}}%
  {\colorbox[#1]{#2}}%
  {#3}\kern-\fboxsep
}
\makeatother
\definecolor{cellgray}{gray}{0.9}
\definecolor{top1-5}{gray}{0.69}
\definecolor{top6-10}{gray}{0.9}
\AtBeginDocument{\setlength{\cmidrulekern}{0.3em}}

\def\BibTeX{{\rm B\kern-.05em{\sc i\kern-.025em b}\kern-.08em
    T\kern-.1667em\lower.7ex\hbox{E}\kern-.125emX}}
\begin{document}

\title{Pull Request Decision Explained: \\An Empirical Overview}

\author{Xunhui Zhang,
        Yue Yu,
        Georgios Gousios,
        and~Ayushi Rastogi 
\IEEEcompsocitemizethanks{\IEEEcompsocthanksitem X. Zhang and Y. Yu are with the National Key Lab of Parallel Distribution, National University of Defense Technology, Changsha, Hunan 410073, China. E-mail: \{zhangxunhui, yuyue\}@nudt.edu.cn.\protect\\
\IEEEcompsocthanksitem A. Rastogi is with the Faulty of Science and Engineering, the University of Groningen, The Netherlands. 
A part of the work was performed while the author was affiliated to TU Delft. 
E-mail: a.rastogi@rug.nl.\protect\\
\IEEEcompsocthanksitem G. Gousios is with Facebook. This work was performed while the author was affiliated with TU Delft. E-mail: gousiosg@fb.com.}}

\markboth{Journal of \LaTeX\ Class Files,~Vol.~14, No.~8, August~2015}%
{Shell \MakeLowercase{\textit{et al.}}: Bare Demo of IEEEtran.cls for Computer Society Journals}

\IEEEtitleabstractindextext{%
\begin{abstract}
\emph{Context}: 
Pull-based development model is widely used in open source, leading the trends in distributed software development.
One aspect which has garnered significant attention is studies on pull request decision - identifying factors for explanation.  
\emph{Objective}: 
This study builds on a decade long research on pull request decision to explain it. 
We empirically investigate how factors influence pull request decision and scenarios that change the influence of factors. 
\emph{Method}: 
We identify factors influencing pull request decision on GitHub through a systematic literature review and infer it by mining archival data.
We collect a total of 3,347,937 pull requests with 95 features from 11,230 diverse projects on GitHub.
Using this data, we explore the relations of the factors to each other and build mixed-effect logistic regression models to empirically explain pull request decision. 
\emph{Results}: 
Our study shows that a small number of factors explain pull request decision with the integrator same or different from the submitter as the most important factor.
We also noted that some factors are important only in special cases
\emph{e.g.,} the percentage of failed builds is important for pull request decision when continuous integration is used.
\end{abstract}

}

\maketitle

\section{Introduction}


\IEEEPARstart{P}{ull-based} development model is an important paradigm for global collaboration in open source projects. 
In this model~\cite{Gousios_exploratory}, contributor \emph{(also known as requester and submitter)} submits their proposed code changes to a base repository by creating a pull request from their cloned repository for reviewers to inspect. 
The integrator \emph{(also known as closer and merger)} evaluates the proposed changes and decide whether to accept or reject the pull request.
This process is, however, made complex with additional actors and mechanisms.
For instance, during the review, anyone can participate to discuss its feature(s), correctness, and more.
Then, there exists Devops tools that automatically check the adaptability of code and give results to contributors and integrators. 

In recent years, many studies on understanding pull-based development have emerged to improve developer contribution, balance integrators' workload, optimize review process, etc.
There are studies on pull request decision~\cite{Iyer_tse}, its latency~\cite{Maddila_latency}, reviewer recommendation~\cite{Jiang_recommendation, Yu-recommendation}, duplication of pull request~\cite{Yu_duplication, Wang_duplication}, automatic generation of pull request description~\cite{Liu_auto_descriptioin}, prioritize pull request lists~\cite{Veen_prioritize}, and others.
This study focuses on explaining the pull request decision.

In the past decade, many studies have made strides at explaining pull request decision by introducing new factors. 
Some examples of these factors are continuous integration~\cite{Yu_determinants, Bogdan_CI}, geographical location~\cite{Rastogi_relationshipESEM}, and bot usage~\cite{Hu_bot, Peng_bot}.
Relatedly, a few studies presented a list of factor that can influence pull request decision. 
Some noteworthy works along this line are the list of developer, project, and pull request characteristics by Gousios et al.~\cite{Gousios_exploratory}.
Another study by Tsay et al.~\cite{Tsay_influence} split factors into two categories, \emph{i.e.} social and technical related factors.
A more recent study by Dey et al.~\cite{Dey_npm} combined many such factors (50) to rank their importance for prediction.

While several studies have contributed individual pieces to understand pull request decision, a systematic synthesis of the body of knowledge to explain pull request decision is missing. 
Along these lines, our current work presents an empirical investigation on explaining pull request decision at GitHub in terms of the factors known to influence it. 
Particularly, we explore:

\begin{enumerate}[start=1,label={\bfseries RQ\arabic*}, leftmargin = 3em]
    \item \emph{How factors influence pull request decisions?}
    \item \emph{Do factors influencing pull request decisions change with context?}
\end{enumerate}

First, we conduct a systematic literature review to identify a comprehensive list of factors known to influence pull request decision.
Then, we create a large and diverse dataset of pull requests and factors (or its indicators) that can be mined from software archival data. 
Finally, we build models (mixed-effect logistic regression) suggesting relations of factors to pull request decisions, in general but also is specific scenarios (e.g., when pull request uses continuous integration).

This paper makes following contributions to software engineering research, practice, and education:
\begin{enumerate} 
    \item We present a curated dataset of 11,230 projects on GitHub with 95 factors and 3,347,937 pull requests. 
    Our dataset is diverse on team size, programming language, and activities (see Table~\ref{project_diversity}). 
It also covers the entire project life cycle as a representation of diversity in time.
\item We present a synthesis of factors identified in literature indicating its significance and direction. 
\item We show the importance of factors in explaining pull request decision and how it changes with context.
\end{enumerate}

The rest of the paper is organized as follows. In Section \ref{study-design}, we explain the design of our research. In Section \ref{results}, we present the results. We discuss the implication in Section~\ref{discussion} and present the threats in Section \ref{threats}. In Section \ref{background}, we describe the related work of this study. In Section \ref{conclusion}, we present conclusions and future work.

\section{Study Design}
\label{study-design}

\begin{figure*}[h!]
  \centering
  \includegraphics[width=7in]{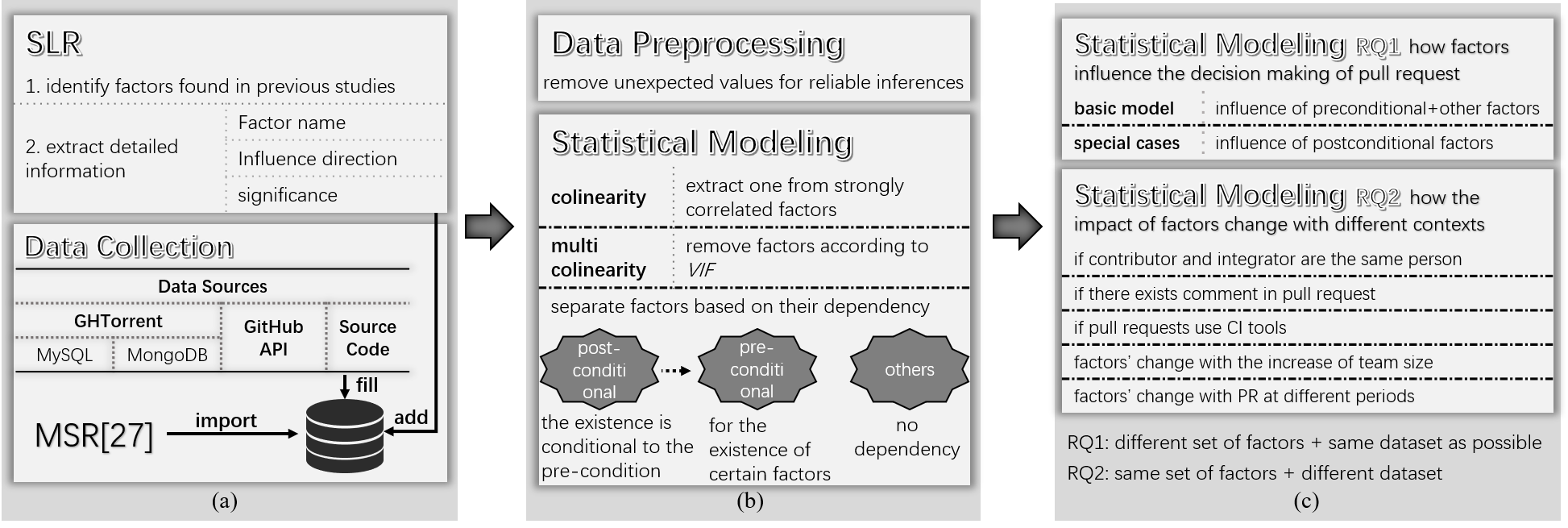}
  \caption{The framework of this paper}
  \label{fig:framework}
\end{figure*}

The framework of our study is shown in Figure~\ref{fig:framework}, which mainly comprises of four parts presenting the steps to empirically explain pull request decision.
First, we gather a comprehensive list of factors known to influence pull request decision (see SLR part in Figure~\ref{fig:framework}.(a)).
Next, we collect data from diverse collaboratively-developed software projects on GitHub as proxies for the factors identified in the above (see Data Collection part in Figure~\ref{fig:framework}.(a)).
Then, we transform the data and bring it to a form usable for analysis (see Data Preprocessing part in Figure~\ref{fig:framework}.(b)). 
Finally, we model the data to answer our research questions, starting with an exploratory data analysis (see Statistical Modeling in Figure~\ref{fig:framework}.(b) and (c)). 

\subsection{Systematic Literature Review}
\label{sub:slr}
To collect all factors known to influence pull request decision, we conducted a systematic literature review (see SLR part in Figure~\ref{fig:framework}.(a)).
Our systematic literature review is based on the guidelines from Kitchenham et al.~\cite{kitchenham_slr}.

Our \emph{search strategy} was to identify all scientific articles relating to pull request decision. 
We selected two widely used terms `pull request' and `pull based' for search that are often used interchangeably as pull request model, pull-based development, and similar other variants. 
We combined the two search terms with a logical `OR' operator (i.e., \emph{"pull request" OR "pull based"}) defining our search space. 

We searched for (\emph{"pull request" OR "pull based"}) in Google Scholar, ACM Digital Library, IEEExplore, Web of Science and EI Compendex,  resulting in a total of 3,941 papers.
We ran the query on 17 April 2020.
We identified 1,000 papers from Google Scholar, 1,433 from ACM Digital Library, 352 from IEEExplore, 487 from Web of Science, and 669 papers from EI Compendex.
We performed an additional step of searching Google Scholar for the papers published only in 2020. 
This step was necessary since Google Scholar retrieves only the top 1,000 results which is likely to miss many articles~\cite{Harzing_pop, Gusenbauer-gs1000}.
The additional search (also conducted on 17 April 2020) resulted in 610 more papers leading to a total of 4,551 papers for snowballing. 

To identify the factors influencing pull request decision, the first author manually analyzed the title and abstract of each paper and selected all studies presenting factors influencing pull request decision that can be inferred by mining software archives.
The search resulted in 19 papers which does not include the papers: 

\begin{itemize}
\item written in languages other than English (45 papers)
\item duplicates (1,181 papers)
\item initial versions of the papers with extended version available (12 papers)
\item presenting factors not applicable to GitHub (5 papers) - \emph{e.g., a study on Firefox and Mozilla core projects shows that `bug severity' and `bug priority' influence acceptance of patch~\cite{Jeong_improving}. 
These attributes do not exist on GitHub.}
\item not related to pull request decision (3,277 papers)
\item related to pull request decision but hard to reproduce (4 papers) - \emph{e.g., using medical equipment to track the eyes of reviewers~\cite{Ford_eye}.}
\item factors not generalizable to a wider range of software projects on GitHub (4 papers) - \emph{e.g., labels~\cite{Pooput_rejection} that vary across communities.}
\item presenting different operationalization of related concepts (3 papers) - \emph{e.g., emotions can be measured directly as joy, love, sadness, and anger; indirectly via valence, arousal, and dominance~\cite{Ortu_mergedIssue}; and abstractly based on polarity~\cite{Iyer_master}.
We choose one of three representations of emotions, i.e., polarity. 
As another example,  Calefato et al.~\cite{Calefato_trust_2017} measured trust using agreeableness - one of the five personality traits used by Iyer et al.~\cite{Iyer_tse}. We choose five personality traits.}
\item presenting factors not measurable quantitatively (1 paper) - \emph{i.e., features relating to pull request decision found in a qualitative study~\cite{Gousios_integrator}.}
\end{itemize}

Next, we identified other relevant articles by navigating through the references of the selected 19 seed articles. 
We applied forward snowballing~\cite{kitchenham_slr} twice, meaning that we selected (a) the references of the 19 articles, and (b) references of the references.
After two rounds, we did not find new related papers.
This process resulted in 7 new papers bringing the total to 26 papers presenting factors relating to pull request decision. 

An overview of the 94 features (factor \emph{same\_user} is not considered in previous studies) found in systematic literature review is shown in Table~\ref{factor_descriptive_statistics}.
Table~\ref{factor_descriptive_statistics} lists the symbolic representation of the features in columns 1 and 3 followed by its description in columns 2 and 4 respectively. 
All the features are classified as developer, project, and pull request characteristics. 
Further, Table~\ref{factor_table} shows the relations of factors to pull request decision, as identified in the 26 selected research articles.

\subsection{Data Collection}
We collected data on a variety of software projects hosted on GitHub as a proxy for the factors identified in the above. 
The dataset used for this study comes from our prior work~\cite{msr2020_ours} featuring 96 factors collected from 11,230 projects.
Further, we enriched the dataset with missing factors and values (see Data Collection part in Figure~\ref{fig:framework}.(a)).

\emph{Our initial dataset}~\cite{msr2020_ours} is built on the publicly available GHTorrent mysql data dump dated 1st June 2019.\footnote{http://ghtorrent-downloads.ewi.tudelft.nl/mysql/mysql-2019-06-01.tar.gz}
It features 96 factors relating to pull request, developer, or project (derived from 76 research articles published in and between 2009 and 2019) for 11,230 software projects. 
The selected projects are source (or base) repositories, excluding forks and deleted repositories.
They are actively developed (with at least one new pull request in the last three months), comes from six programming languages and have different size and activity levels (see Table~\ref{project_diversity} for an overview).
For the selected 11,230 projects, the dataset offers a total of 3,347,937 closed pull requests (meaning a decision has been made) that are submitted to the default branch of the repository.

\begin{table}[htbp]
\centering
\scriptsize
\renewcommand{\arraystretch}{1.37}
\caption{The description of project diversity}
\label{project_diversity}
\begin{tabular}{*5l}    \toprule
\textbf{category} & \textbf{type} & \textbf{project count} & \textbf{percentage} \\\midrule
\multirow{6}{*}{language} & JavaScript & 3,879 & 34.5\% \\ 
\cdashline{2-4}[0.8pt/2pt]
 & Python & 3,055 & 27.2\% \\ 
 \cdashline{2-4}[0.8pt/2pt]
 & Java & 1,823 & 16.2\% \\
 \cdashline{2-4}[0.8pt/2pt] 
 & Ruby & 1,243 & 11.1\% \\
 \cdashline{2-4}[0.8pt/2pt] 
 & Go & 913 & 8.1\% \\ 
 \cdashline{2-4}[0.8pt/2pt]
 & Scala & 317 & 2.8\% \\ 
\hline
\multirow{3}{*}{project size} & small \emph{$\leq 12$ developers} & \multirow{1}{*}{3,711} & \multirow{1}{*}{33\%} \\
 \cdashline{2-4}[0.8pt/2pt]
 & mid \emph{$\leq 31$ developers} & \multirow{1}{*}{3,634} & \multirow{1}{*}{32.4\%} \\
 \cdashline{2-4}[0.8pt/2pt]
 & large \emph{$>31$ developers} & \multirow{1}{*}{3,885} & \multirow{1}{*}{34.6\%} \\
\hline
\multirow{5}{*}{project activity} & min \emph{$=33$ pull requests} & \multirow{1}{*}{-} & \multirow{1}{*}{-} \\
 \cdashline{2-4}[0.8pt/2pt]
 & 25\% \emph{$\leq 55$ pull requests} & \multirow{1}{*}{2,843} & \multirow{1}{*}{25.3\%} \\
 \cdashline{2-4}[0.8pt/2pt]
 & 50\% \emph{$\leq 106$ pull requests} & \multirow{1}{*}{2,796} & \multirow{1}{*}{24.9\%} \\
 \cdashline{2-4}[0.8pt/2pt]
 & 75\% \emph{$\leq 261$ pull requests} & \multirow{1}{*}{2,791} & \multirow{1}{*}{24.9\%} \\
 \cdashline{2-4}[0.8pt/2pt]
 & max \emph{$=38,953$ pull requests} & \multirow{1}{*}{-} & \multirow{1}{*}{-} \\
\bottomrule
\hline
\end{tabular}
\end{table}

Our initial dataset is  futuristic and emphasizes generalizability - a design choice for a wide range of explorations~\cite{msr2020_ours}. 
It has 12 times more projects and 10 times more pull requests in comparison to Gousios et al’s dataset~\cite{Gousios_Dataset}.
It is more diverse than any prior study on GitHub focusing on pull request decision which have largely focused on the most popular projects.

From Table~\ref{project_diversity}, we can see that the diversity of selected projects is mainly manifested in three aspects, \emph{i.e.} covering 6 languages, containing different team sizes and including projects with different activity levels (the number of pull requests ranges from 33 to more than 30 thousand).
Its features are applicable to projects outside GitHub and has additional features which are likely to influence pull request development- an extrapolation of existing features.

For our analysis, we selected data relating to the factors identified by our systematic literature review from the initial dataset.
We noticed that some of the factors identified by our systematic literature review do not exist in the initial dataset, so we added the missing features. 
Table~\ref{factor_descriptive_statistics} presents a complete list of factors known to influence pull request decision on GitHub. 
Factors marked as $\star$ are additions of our study to the initial dataset~\cite{msr2020_ours}. 
\begin{table*}[htbp]
\centering
\scriptsize
\caption{presents a comprehensive list of factors known to influence pull request decision on Github.}
\label{factor_descriptive_statistics}
\begin{tabular}{| p{2.7cm} | p{5.3cm} | p{3.3cm} | p{5.4cm} |}
\toprule
Factor & Description & Factor & Description \\  \hline
\multicolumn{4}{|c|}{\textit{\textbf{Developer Characteristics}}}  \\ \hline
first\_pr   & first pull request? - yes/no & prior\_review\_num   & \# previous reviews in a project\\
core\_member    & core member? - yes/no &first\_response\_time & \# minutes from pull request creation to reviewer's first response\\
contrib\_gender   & gender - male and female & contrib\_country & country of residence\\
same\_country & same country contributor/integrator? - yes/no & prior\_interaction & \# interactions to a project in the last three months \\
same\_affiliation & same affiliation contributor/integrator? - yes/no & contrib/inte\_affiliation & contributor/integrator affiliation\\
contrib/inte\_\textbf{X}   & contributor/integrator personality traits (\emph{open}: openness; \emph{cons}: conscientious; \emph{extra}: extraversion; \emph{agree}: agreeableness; \emph{neur}: neuroticism) & perc\_contrib/inte\_X\_emo & \% contributor/integrator (\emph{neg}: negative/\emph{pos}: positive) emotion in comments\\
X\_diff & absolute difference in the personality traits of contributor and integrator & contrib/inte\_first\_emo& emotion in contributor/integrator's first comment \\
social\_strength & fraction of team members interacted in last three months & contrib\_follow\_integrator & contributor followed integrator before pull request creation? yes/no\\ 
followers   & \# followers at pull request creation time & same\_user $\bullet$   & same contributor and integrator? - yes/no \\
prev\_pullreqs   & \# previous pull requests & account\_creation\_days   & \# days from contributor's account creation to pull request creation \\
contrib\_perc\_commit $\star$ & \% of previous contributor's commit & & \\
\hline
\multicolumn{4}{|c|}{\textit{\textbf{Project Characteristics}}}\\ \hline
sloc   & executable lines of code & team\_size   & \# active core team members in last three months \\
language  & programming language & open\_issue\_num   & \# open issues\\
project\_age   & \# months from project to pull request creation & open\_pr\_num   & \# open pull requests\\
pushed\_delta   & \# seconds between two latest pull requests open & fork\_num & \# forks\\ 
pr\_succ\_rate   & pull request acceptance rate of project & test\_lines\_per\_kloc   & \# test lines per 1K lines of code \\ 

stars   & \# stars & integrator\_availability   $\star$ & latest activity of the two most active integrators\\
test\_cases\_per\_kloc & \# test cases per 1K lines of code & asserts\_per\_kloc & \# assertions per 1K lines of code \\
perc\_external\_contribs & \% external pull request contributions & requester\_succ\_rate & past pull request success rate \\
\hline
\multicolumn{4}{|c|}{\textit{\textbf{Pull Request Characteristics}}}  \\ \hline
churn\_addition & \# added lines of code & churn\_deletion& \# deleted lines of code \\
bug\_fix & fixes a bug? - yes/no & description\_length   & length of pull request description\\
test\_inclusion   & test case exists? - yes/no & comment\_conflict   & keyword `conflict' exists in comments? - yes/no\\
hash\_tag    & \# tag exists? & num\_participants & \# participants in pull request comments \\
lifetime\_minutes  & \# minutes from pull request creation to lastest close time & part\_num\_code & \# participants in pull request and commit comments \\
ci\_exists   & uses continuous integration? - yes/no & ci\_build\_num & \# CI builds\\
ci\_latency & \# minutes from pull request creation to the first CI build finish time & perc\_neg\_emotion & \% negative emotion in comments\\
num\_code\_comments $\star$ & \# code comments & perc\_pos\_emotion & \% positive emotion in comments \\
test\_churn   & \# lines of test code changed (added + deleted) & num\_code\_comments\_con $\star$ & \# contributor's code comments\\
ci\_test\_passed   & all CI builds passed? - yes/no & ci\_first\_build\_status & CI first build result\\
ci\_failed\_perc & \% CI build failed & ci\_last\_build\_status\ & CI last build status\\
num\_commits   & \# commits & src\_churn   & \# lines changed (added + deleted)\\
files\_added   & \# files added & files\_deleted   & \# files deleted\\
files\_changed   & \# files touched & friday\_effect   $\star$ & pull request submitted on Friday? -yes/no \\
reopen\_or\_not  $\star$ & pull request is reopened? - yes/no & commits\_on\_files\_touched   & \# commits on files touched  \\
has\_comments   $\star$ & pull request has comment? -yes/no & num\_comments   & \# comments \\
has\_participants $\star$ & has participant? -yes/no & core\_comment $\star$ & has core member comment? -yes/no \\
contrib\_comment $\star$ & has contributor comment? -yes/no & inte\_comment $\star$ & has integrator comment? -yes/no \\
has\_exchange $\star$ & has contributor/integrator comment? -yes/no & other\_comment   $\star$ & non-contributor/core team comment? -yes/no \\
num\_comments\_con $\star$ & \# contributor comments & at\_tag & @ tag exists? -yes/no \\
\bottomrule
\multicolumn{4}{l}{NOTE: Factors marked as $\star$, are additions of our study to the latest MSR Data Showcase pull request dataset~\cite{msr2020_ours}, while $\bullet$ are additions to previous studies.} \\
\multicolumn{4}{l}{\emph{All metrics are relative to a referenced pull request in a project.}} \\
\multicolumn{4}{l}{\emph{Factors that change over time (e.g., core team) are measured using the previous three months of development activities in a project.}} \\

\end{tabular}
\end{table*}

Finally, we enriched our dataset by filling missing values where possible based on GHTorrent\footnote{https://ghtorrent.org/}, GitHub API and source code of repository.
For example, the initial dataset used the tool by Vasilescu et al.~\cite{Bogdan_gender_tool} to infer country information. 
The resulting dataset, however, had a large number of missing values.
We applied several steps such as using \emph{country\_code} information and \emph{pycountry} package\footnote{https://pypi.org/project/pycountry/} to generate country name. 
This way we were able to derive country information of additional 546,682 contributors (1,473,008 before), 747,204 integrators (1,580,256 before) and 796,083 same country participants (1,081,668 before).
The expanded country information can be seen on Github.\footnote{https://github.com/zhangxunhui/TSE\_pull-based-development/blob/master/country\_info.csv}

We added a factor \emph{same\_user} which do not exist in prior studies (marked as $\bullet$ in Table~\ref{factor_descriptive_statistics}). 
While information on same user is not useful in itself, it adds meaning to factors such as \emph{same\_country}, \emph{same\_affiliation} and personality difference related factors (\emph{e.g. open\_diff}) which only make sense when the contributor and integrator are not the same user.
In our dataset, we found that 43.6\% of the pull requests are integrated by the submitters (85.7\% of them are core contributors and 14.3\% are external contributors). Comparing to directly committing to code repository, pull-based development is becoming a standard collaborative mode where not only external contributors but also core members are following. Therefore, it is necessary to add this factor and study its influence on pull request decision. 

\subsection{Data Preprocessing}
Our exploration of the resulting dataset (manually and using data distribution graphs) showed some unexpected data values for factors such as \emph{first\_response\_time}, \emph{ci\_latency}, \emph{account\_creation\_days} and \emph{project\_age}. 
While several factors can contribute to the problem (such as metric definition and missing data), it is important to fix them for reliable inferences (see Technical Report~\cite{Zhang-msr-report} for examples). 


\begin{itemize}
    \item \emph{first\_response\_time} 
    has \emph{negative} values for some pull requests. 
    This happens since our metric not only considers the discussion under a pull request, but also the comments under the related code. 
    Since some comments exist before pull request creation, our data shows negative values. 
    We fix it by removing pull requests with negative values (0.4\%).
    
    \item \emph{ci\_latency} has \emph{negative} values for some pull requests. 
    CI latency measures the time from pull request creation until CI built finish time. 
    In some cases, however, commit(s) exist prior to pull request creation and the time of first build recorded is earlier than the creation time of a pull request. 
     We fix the problem by removing those pull requests (1.5\%).
     
    \item \emph{account\_creation\_days} and \emph{project\_age} have \emph{negative} values.
   It happens in special cases where the creation time of a user account in GHTorrent is different from Github API. 
    Here too, we remove the entry (0.1\%).
    
  \item \emph{bug\_fix} has 99.3\% empty values.
  We remove this factor which otherwise can adversely affect the analysis.
\end{itemize}

\subsection{Statistical Modeling}
Presenting comprehensive analysis of the factors influencing pull request decision,
we build generic models, comprising of all the factors, and models representing specific cases. 
We also build models within different contexts.
But first, we explore relationships among the factors identified in the above.

Our preliminary exploration into the relationship among factors started with calculating correlations among all the factors.
We calculate Spearman correlation coefficient ($\rho$) for continuous factors~\cite{Gousios_exploratory}, 
Cramér's V ($\Phi_c$) for categorical factors~\cite{StatsTest-cramer}, and partial Eta-squared ($\eta^2$) for correlation between continuous and categorical factors~\cite{Jones-eta}. 
We refer to $\rho>0.7$~\cite{Gousios_exploratory}, $\Phi_c>\frac{0.5}{df}$~\cite{cohen_1969}, and $\eta^2>0.14$~\cite{cohen_1969} as strong correlations.

A list of strongly correlated factors is presented in Table~\ref{strongly_correlated_factors}.
In Table~\ref{strongly_correlated_factors}, strongly correlated factors are separated from the other factors by a dotted line.
For a complete list of correlation between each pair of factors, refer to our technical report~\cite{Zhang-msr-report}.
\begin{table}[htbp]
\centering
\caption{presents strongly correlated factors among the factors known to influence pull request decision on GitHub. 
We choose one of the correlated factors to build statistical model.
The reason for the choice is presented below.}
\label{strongly_correlated_factors}
\begin{tabular}{ p{3cm} p{2.5cm} p{1.8cm} }
  \toprule
  \textbf{Correlated factors} & \textbf{Selected factor} & \textbf{Reason} \\ \hline
  
  test\_lines\_per\_kloc & \multirow{3}{*}{\emph{test\_lines\_per\_kloc}} & \multirow{3}{*}{previous study}\\
  test\_cases\_per\_kloc & & \\
  asserts\_per\_kloc & &  \\ \hdashline
  
  src\_churn & \multirow{3}{*}{\emph{src\_churn}} & \multirow{3}{*}{frequency} \\
  churn\_addition & & \\
  churn\_deletion & & \\ \hdashline

  num\_comments & \multirow{4}{*}{\emph{num\_comments}} & \multirow{4}{*}{frequency} \\
  at\_tag & & \\
  num\_participants & & \\
  num\_comments\_con & & \\ \hdashline
  
  core\_member & \multirow{4}{*}{\emph{core\_member}} & \multirow{4}{*}{frequency} \\
  perc\_external\_contribs & & \\
  social\_strength & & \\
  requester\_succ\_rate & & \\ \hdashline
  
  stars & \multirow{3}{*}{\emph{stars}} & \multirow{3}{*}{frequency} \\
  fork\_num & & \\
  inte\_affiliation & & \\ \hdashline
  
  prev\_pullreqs & \multirow{2}{*}{\emph{prev\_pullreqs}} & \multirow{2}{*}{frequency} \\
  prior\_interaction & \\ \hdashline
  
  num\_code\_comments & \multirow{3}{*}{\emph{num\_code\_comments}} & \multirow{3}{*}{frequency} \\
  part\_num\_code & & \\
  num\_code\_comments\_con & & \\ \hdashline
  
  open\_pr\_num & \multirow{2}{*}{\emph{open\_pr\_num}} & \multirow{2}{*}{frequency} \\
  fork\_num & \\ \hdashline

  ci\_latency & \multirow{2}{*}{\emph{ci\_latency}} & \multirow{2}{*}{importance} \\
  ci\_build\_num & & \\ \hdashline

  sloc & \multirow{2}{*}{\emph{sloc}} & \multirow{2}{*}{importance} \\
  language & & \\ \hdashline

  has\_comments & \multirow{6}{*}{\emph{has\_comments}} & \multirow{6}{*}{expressiveness} \\
  has\_participants & & \\ 
  core\_comment & & \\
  contrib\_comment & & \\
  inte\_comment & & \\
  has\_exchange & & \\ \hdashline
  
  prior\_review\_num & \multirow{2}{*}{\emph{prior\_review\_num}} & \multirow{2}{*}{data availability} \\
  inte\_affiliation & \\ \hdashline

  open\_issue\_num & \multirow{2}{*}{\emph{open\_issue\_num}} & \multirow{2}{*}{data availability} \\ 
  inte\_affiliation & \\ \hdashline
  
  inte\_cons & \multirow{2}{*}{\emph{inte\_cons}} & \multirow{2}{*}{data availability} \\
  inte\_affiliation & \\ \hdashline
  
  inte\_extra & \multirow{2}{*}{\emph{inte\_extra}} & \multirow{2}{*}{data availability} \\
  inte\_affiliation & \\ \hdashline
  
  inte\_agree & \multirow{2}{*}{\emph{inte\_agree}} & \multirow{2}{*}{data availability} \\
  inte\_affiliation & \\ \hdashline

  same\_country & \multirow{2}{*}{\emph{same\_country}} & \multirow{2}{*}{discussion} \\
  contrib\_country & & \\ \hdashline
  
  perc\_contrib\_pos\_emo & \multirow{2}{*}{\emph{perc\_contrib\_pos\_emo}} & \multirow{2}{*}{discussion} \\
  contrib\_first\_emo & \\ \hdashline
  
  perc\_inte\_neg\_emo & \multirow{2}{*}{\emph{perc\_inte\_neg\_emo}} & \multirow{2}{*}{discussion} \\
  inte\_first\_emo & & \\ \hdashline
  
  perc\_inte\_pos\_emo & \multirow{2}{*}{\emph{perc\_inte\_pos\_emo}} & \multirow{2}{*}{discussion} \\
  inte\_first\_emo & & \\ \hdashline
  
  same\_user & \multirow{5}{*}{\emph{same\_user}} & \multirow{5}{*}{discussion} \\
  inte\_first\_emo & \\ 
  inte\_affiliation & \\
  contrib\_affiliation & \\ 
  contrib\_rate\_author & \\ \hdashline

  same\_affiliation & \multirow{2}{*}{\emph{same\_affiliation}} & \multirow{2}{*}{discussion} \\
  contrib\_affiliation & & \\ \hdashline

  perc\_neg\_emotion & \multirow{4}{*}{\emph{perc\_neg\_emotion}} & \multirow{4}{*}{discussion} \\
  perc\_contrib\_neg\_emo & & \\
  contrib\_first\_emo & & \\
  inte\_first\_emo & & \\ \hdashline
  
  perc\_pos\_emotion & \multirow{2}{*}{\emph{perc\_pos\_emotion}} & \multirow{2}{*}{discussion} \\
  inte\_first\_emo & & \\ \hdashline

  ci\_failed\_perc & \multirow{4}{*}{\emph{ci\_failed\_perc}} & \multirow{4}{*}{discussion} \\
  ci\_test\_passed & & \\
  ci\_first\_build\_status & & \\
  ci\_last\_build\_status & & \\ \bottomrule
  
\end{tabular}
\end{table}

Next, we build mixed-effects logistic regression model(s) to empirically explain factors influencing the pull request decision. 
The models use project identifier as a random effect, indicating similarity among the pull requests from a project~\cite{linear_mixed_effect}. 
All other factors have fixed effects.
The resulting model indicates the significance of a factor and direction of its association to a pull request decision - accept or reject.
We use \emph{glmer} function of \emph{lme4}~\cite{R_lme4} package in R to model the pull request decision.

To build model for explanation, we include all factors that can be meaningfully added together, do not present same or similar information as other factors, and are easy to interpret. 

\begin{enumerate}
\item \emph{Adding meaningful factors} While adding factors to a model, we observed that some factors do not make sense outside a specific context.
For example, if contributor and integrator is the same, factors such as `personality difference' do not exist and make no sense.
We refer to such factors as `pre-conditional factors' and `post-conditional factors'.
\emph{`Pre-conditional factors'} are those which must exist for another factor to exist and make sense (e.g., \emph{same\_user} in the previous example). 
On the other hand, \emph{`post-conditional factors'} are the factors whose existence is conditional to the pre-conditional factors (e.g., \emph{open\_diff}). 
All the other factors are classified to `others' category.
A complete list of pre and post-conditional factors is presented in Table~\ref{factor_dependent_table}. 
\item \emph{Factors presenting same information} Our preliminary investigation shows that several factors identified from the literature are strongly correlated to each other (see Table~\ref{strongly_correlated_factors} for a list of strongly correlated factors).
When two related factors are added to a model, the factor not only changes the pull request decision but also the other factor. 
This can change the estimated effect of a factor on the pull request decision and its significance, also referred to as multi-collinearity problem~\cite{multicollinearity-problem}. 
To avoid multi-collinearity, we select one of the many strongly correlated factors.
Our choice for the selection of a factor was influenced by its use in previous studies (e.g., \cite{Gousios_exploratory} chose \emph{test\_lines\_per\_kloc}), frequency of occurrence in literature (e.g., \emph{core\_member} appeared most often), importance (e.g., \emph{sloc} influences pull request decision), expressiveness (e.g., \emph{has\_comments} is broader and more informative than \emph{contrib\_comment}), data availability (e.g., \emph{open\_issue\_num} has most non-empty values), and otherwise in discussion with the last author (e.g., \emph{perc\_pos\_emotion}). 
We also exclude factors with VIF value $\geq$ 5 that can inflate variance, measured using \emph{vif} function of \emph{car} package in R~\cite{cohen_applied}.
This way we removed \emph{num\_code\_comments} that is otherwise moderately correlated with \emph{num\_comments} ($\rho=0.63$).

\item \emph{Ease of interpretation} Models perform better when features are approximately normal and on comparable scale.\footnote{https://medium.com/@sjacks/feature-transformation-21282d1a3215}
We stabilize the variance in features by adding a value `1' and log-transforming continuous variables. 
Later, we transform the features to a comparable scale with mean value `0' and standard deviation `1'.
\end{enumerate}
\begin{table}[htbp]
\centering
\caption{presents factors with dependency. Post-conditional factors are dependent on the pre-conditional factors for existence and meaning.}
\label{factor_dependent_table}
\begin{tabular}{ p{3cm} p{3cm} }
  \toprule
  \textbf{post-conditional factor} & \textbf{pre-conditional factor} \\ \hline
  
  \emph{perc\_pos\_emotion} & \multirow{3}{*}{\emph{has\_comments}} \\
  \emph{perc\_neg\_emotion} &  \\ 
  \emph{first\_response\_time} & \\ \hdashline
  
  \emph{perc\_contrib\_pos\_emo} & \multirow{2}{*}{\emph{contrib\_comment}} \\
  \emph{perc\_contrib\_neg\_emo} & \\ \hdashline
  
  \emph{perc\_inte\_neg\_emo} & \multirow{2}{*}{\emph{inte\_comment}} \\ 
  \emph{perc\_inte\_pos\_emo} & \\ \hdashline
  
  \emph{ci\_latency} & \multirow{2}{*}{\emph{ci\_exists}} \\
  \emph{ci\_failed\_perc} & \\ \hdashline
  
  \emph{same\_country} & \multirow{8}{*}{\emph{same\_user}} \\ 
  \emph{same\_affiliation} & \\
  \emph{contrib\_follow\_integrator} & \\
  \emph{open\_diff} & \\
  \emph{cons\_diff} & \\
  \emph{extra\_diff} & \\
  \emph{agree\_diff} & \\
  \emph{neur\_diff} & \\ 
  \bottomrule
  
\end{tabular}
\end{table}

\subsubsection{Factors influencing pull request decision}
To explain pull request decision, we intended to build a model with all the known factors.
In practice, this is not possible.
We noticed that some factors do not make sense unless a pre-condition is met.
For example, factor \emph{ci\_latency} is meaningful only when factor \emph{ci\_exists} is true.
Here, \emph{ci\_exists} presents a pre-condition contingent on which factors such as \emph{ci\_latency} are meaningful, also referred to as post-conditional factors.
Table~\ref{factor_dependent_table} presents a complete list of dependent factors in our dataset.
The remaining factors have no such dependency to other factors (Figure~\ref{fig:framework}.(b) presents an overview).

To understand how factors influence pull request decision, we build two types of models. 
\begin{enumerate}
\item We build a \emph{basic model} which comprises of all the factors with no dependencies on each other (referred as `others' in Figure~\ref{fig:framework}.(b)) and pre-conditional factors.
This model offers an overview without entering into the details offered by the post-conditional factors.
\item Next, we build models for the \emph{special cases} relating to pre-conditions.
We build models relating to developer, pull request, and tools for the pre-conditions identified in Table~\ref{factor_dependent_table}.
\begin{itemize}
    \item \emph{developer:} when contributor and integrator are not the same user (\emph{same\_user=0})
    \item \emph{pull request:} when a pull request has comments (\emph{has\_comment=1})
    \item \emph{tool:} when a pull request uses CI tool (\emph{ci\_exists=1})
    Each of these special case models are build on a subset of the data used in the basic model that meets the pre-condition.  
  \end{itemize}
\end{enumerate}

\subsubsection{Influence of context}
\label{additionals-RQ2}
\begin{figure}[h!]
  \centering
  \includegraphics[width=3in]{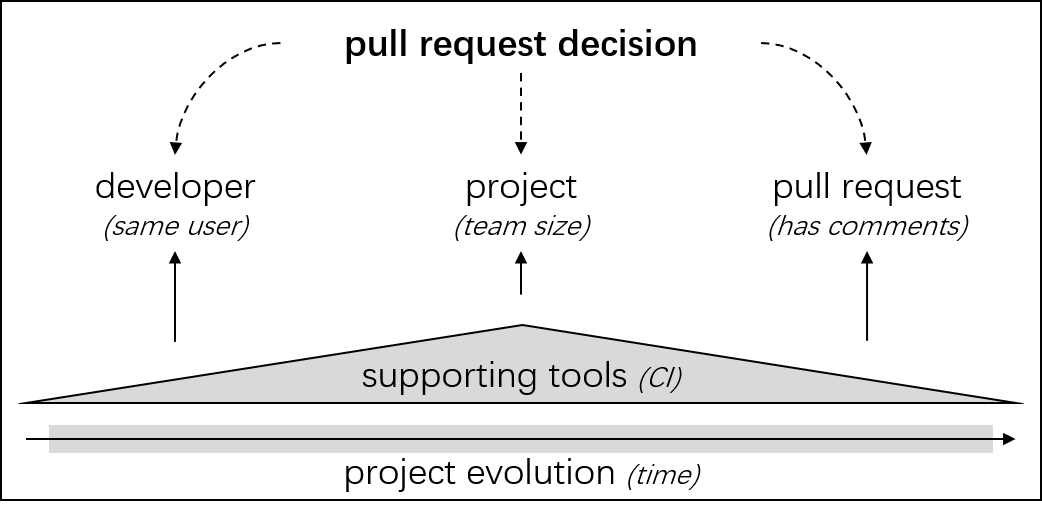}
  \caption{presents the contexts in pull request decision}
  \label{fig:diff-contexts-framework}
\end{figure}

To explore the relevance of context in explaining pull request decision, we study five scenarios relating to developer, pull request, project, tools, and time.
Figure~\ref{fig:diff-contexts-framework} presents a pictorial depiction of the five scenarios in relation to the pull request decision and metrics. 
To study the influence of context, we train the same model on different observations representing specific contexts.

\begin{itemize}
    \item \emph{developer characteristic:} 
    We choose factor \emph{same\_user} indicating whether a pull request is submitted and integrated by the same user.
    It is the most important developer characteristic influencing pull request decision (see basic model in Table~\ref{result-special-case}) and also a pre-condition for a range of factors.
    We believe that the pull requests integrated by self behave differently than the pull requests integrated by others.

    \item \emph{pull request characteristic:} 
    We choose factor \emph{has\_comments} as an indicator of pull request characteristic influencing the decision~\cite{Tsay_influence}.
    It is one of the top five factors influencing decision (see basic model in Table~\ref{result-special-case}) and a pre-condition for several factors including \emph{perc\_pos\_emotion} and \emph{first\_response\_time} (see Table~\ref{factor_dependent_table}).
    This factor explores decisions for pull requests with and without comments.

    \item \emph{project characteristic:} 
    We selected factor \emph{team\_size} as an indicator of project characteristics such as project popularity and maturity.
    We assume that different sized teams represented different contexts (as is also seen in other studies~\cite{Pcndharkar-team-size, Chou-team-size}). 
    We study three team sizes: small ($\leq 4$ contributors), medium ($4 < team\_size \leq 10$), and large ($team\_size > 10$). 
    
    \item \emph{supporting tools:} 
   We select factor \emph{ci\_exists} for its reported influence on pull request decision~\cite{Yu_determinants} and relevance in our special case model (referring to Table~\ref{result-special-case}).
   Not only this, a study shows that the usage of CI tools change during the development of project~\cite{Zhao_ci}.
   Therefore, we assume that factors influence pull request decision differently between pull requests using CI tools and not using CI tools. 
    
    \item \emph{project evolution:} 
    We study temporal evolution to see if the process changes over time. 
    We study decision making in three time periods namely before June 1st 2016, between Jun 1st 2016 and Jun 1st 2018, and between Jun 1st 2018 and Jun 1st 2019.
    A pull request belongs to a time period when it is integrated. 
    For this scenario, we only include projects (and its pull requests) active in all three time periods.
\end{itemize}

\subsubsection{Interpreting Statistical Models}
The resulting mixed-effects logistic regression models indicates the influence of a factor in a model and its relative relevance.
Section~\ref{results} presents the findings from mixed-effect logistic regression models.
Each model has two parts: an intercept and influence of a factor expressed as:

\begin{equation}
  \label{result-equation}
odds\;ratio^{p-value}[percentage\;variance]
\end{equation}

\emph{Odds ratio} expresses the association between a factor and pull request decision as ``the increase or decrease in the odds of acceptance for a `unit' increase of a factor''~\cite{Tsay_influence}.
Here, a `unit' of each factor is one standard deviation from the standardization of log-transformed factors.
The term \emph{p-value} indicates the statistical significance of a factor, and significance is indicated by stars: *** p<0.001; ** p<0.01; * p<0.05.
Finally, \emph{percentage of explained variance} is used as a proxy for the relative importance of a factor.
This metric is similar to the percentage of total variance explained by least square regression~\cite{cohen_applied} and is used in prior studies~\cite{perc-variance}.

We report the \emph{goodness of fit} for each model using AUC value (for training data) where $AUC>0.5$ indicates the effectiveness of a model~\cite{Rastogi_relationshipESEM}.
We also report the predictive performance of related models using weighted precision, weighted recall, and weighted f-score\cite{Bradley_AUC}.

In practice, we split the pull requests on close time, and use the first 90\% of pull requests for training and the remaining 10\% for testing.
We measure the predictive performance for basic model only to present the prediction effect of pull request decision by integrating as many factors as possible, and explain factors' performance in other situations without reporting their performance in prediction.
The above metrics collectively indicates the predictive performance of both baseline and logistic regression model for our highly imbalanced dataset~\cite{Bradley_AUC}.

\section{Results}
\label{results}
This section presents how factors influence pull request decisions (answering \textbf{RQ1}) via a basic model.
The basic model comprises of all the factors likely to influence pull request decisions, excluding the factors which cannot make it to the basic model. 
Next, we describe how the factors influencing pull request decisions change with context (answer to \textbf{RQ2}).
We present five scenarios representing developer, pull request, project, tool, and time characteristics. 

\subsection{RQ1: How factors influence pull request decisions?}

\subsubsection{Basic model}
\label{decision_making_whole}

Our basic model in Table~\ref{result-special-case} (column 3) shows 46 factors known to influence pull request decision arranged in the non-increasing order of relative relevance.
In comparison to a random classifier (with weighted precision: 0.81, weighted recall: 0.79, weighted f-score: 0.80, and AUC\_test: 0.50), our basic model outperforms (with weighted precision: 0.89, weighted recall: 0.90, weighted f-score: 0.89, and AUC\_test: 0.82) suggesting an improvement for decision making.

The five most important factors influencing pull request decisions are \emph{same\_user}, \emph{lifetime\_minutes}, \emph{prior\_review\_num}, \emph{has\_comments} and \emph{core\_member}.
Table~\ref{result-special-case} (column 3) shows that the top five factors (shown as dark gray) explains approximately 83\% of the variance explained by the model.
This number goes up to approximately 95\% when considering the influence of the top 10 factors.
The remaining 36 factors collectively explain 5\% of the explained variance.

The most important factor influencing the pull request decision is \emph{same\_user} (with 31\% variance).
Factor \emph{same\_user} decreases the odds of acceptance of a pull request by 47.6\% per unit when a pull request is integrated by the contributor.
One possible explanation for the observation relates to the process of pull-based development.
 Due to the standardized process of pull-based development, contribution should be reviewed and merged by others during the process. However, since all contributors can close their own pull requests, it is possible for contributors to find problems in their pull requests from others' comments and close their pull requests during the review process.

\begin{figure*}[h!]
    \centering
    \includegraphics[width=6.5in]{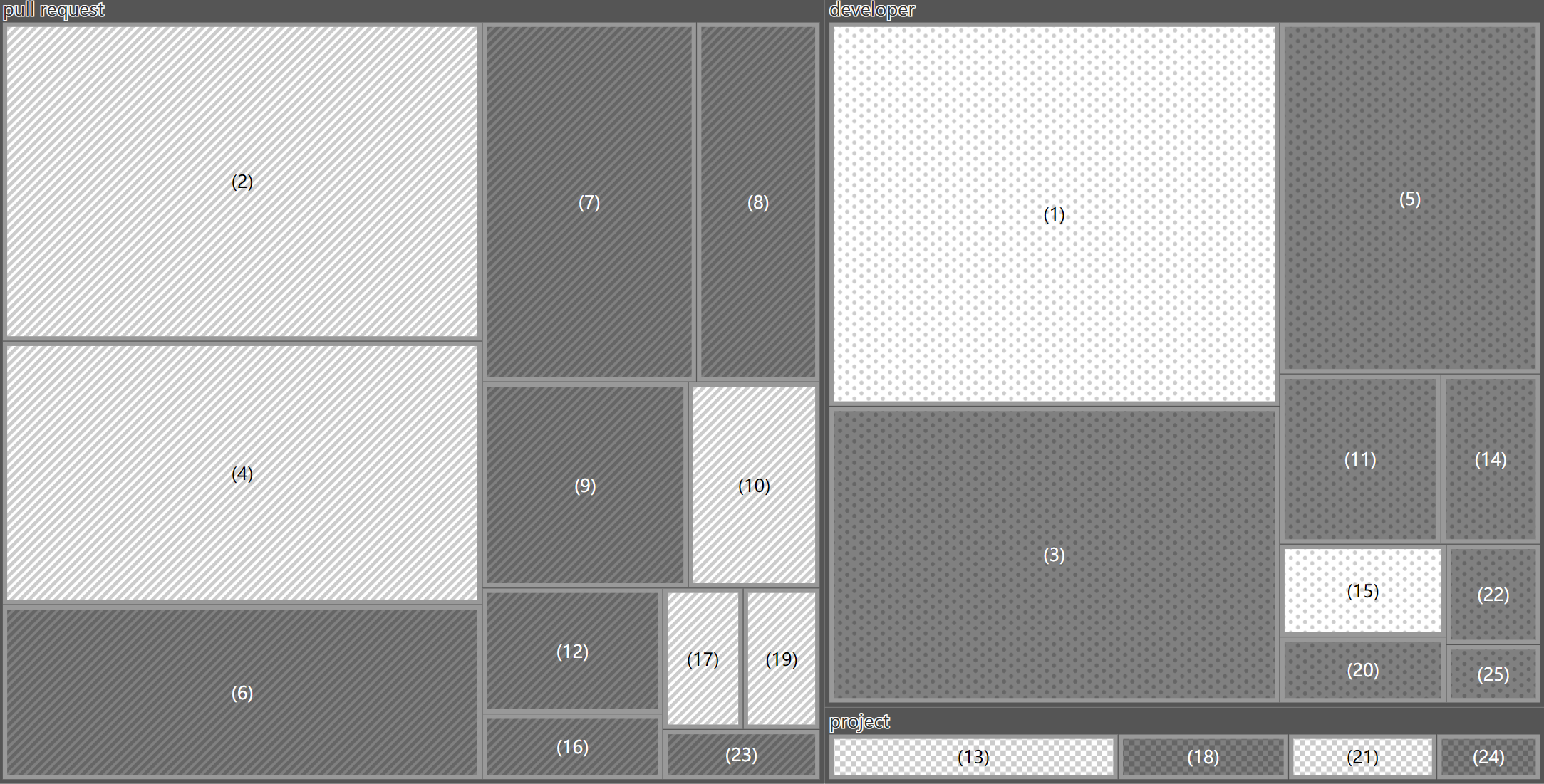}
    \caption{presents the result of basic model. The size of rectangle represents the importance of factors, where it is calculated by the \emph{log} value of percentage of explained variance (the bigger the more important).
    Color means the influence direction (gray means positive and white means negative).
    Text in square represents the factor index, which is shown in Table~\ref{result-special-case}.
    Here we only present factors which explain more than 0.1\% of variance for visualization.}
    \label{fig:circle-result-basic}
  \end{figure*}

Figure~\ref{fig:circle-result-basic} presents an overview of the relevance and directional influence of factors on pull request decision,
when adding up the effect size of different categories (developer, pull request and project).
 It is surprising to find that the project-related factors do not contribute a lot to the decision making of pull request, by only explaining about 1\% of the variance.
 This means that for project-related proxies indicating workload (\emph{open\_pr\_num}), openness (\emph{open\_issue\_num}), etc., do not play a decisive role in the decision to merge a pull request.
Relatively, the developer and pull request related factors are more important, explaining 52\% and 46\% variance respectively.

\subsubsection{Special cases}
\label{special-cases}
Table~\ref{result-special-case} shows the result of the three special cases in the last three columns.
Factors ranking top 5 in each model ($T_{1-5}$) are shown in deep gray, and factors ranking top 6-10 in each model ($T_{6-10}$) are shown in light gray.
\begin{table*}[!h] \centering
    \setlength\tabcolsep{13pt}
    \renewcommand\arraystretch{1.12}
    \scriptsize
    \caption{presents the results of special cases. - means the factor is not included in the model.
    Color: {\color{gray!110}{deep gray}} represents factors with explained variance ranks in Top 5 and {\color{gray!70}{light gray}} represents factors ranks from the 6th to the 10th.} 
    \label{result-special-case} 
  \begin{tabular}{ clcccc} 
  \\[-3ex]
  \toprule
  \\[-3ex] 
  Factor Index & \multicolumn{5}{c}{\textit{Dependent variable: merged\_or\_not=1}} \\ 
  \cline{1-6} 
  \\[-3ex] 
  & & \emph{basic model} & \emph{same\_user=0} & \emph{has\_comments=1} & \emph{ci\_exists=1}\\ [-0.5ex]
  \cmidrule(lr){3-3} \cmidrule(lr){4-4} \cmidrule(lr){5-5} \cmidrule(lr){6-6} \\[-3.3ex] 
    & (Intercept) & $21.1^{***}$ & $32.5^{***}$ & $15.4^{***}$ & $26.1^{***}$ \\ 
    
   (1) & same\_user & \ccell{top1-5}{$0.52^{***}[31.17]$} & - & \ccell{top1-5}{$0.60^{***}[21.53]$} & \ccell{top1-5}{$0.50^{***}[21.27]$}  \\
   (2) & lifetime\_minutes & \ccell{top1-5}{$0.61^{***}[21.10]$} & \ccell{top1-5}{$0.52^{***}[43.08]$} & \ccell{top1-5}{$0.53^{***}[30.40]$} & \ccell{top1-5}{$0.50^{***}[26.08]$}  \\
   (3) & prior\_review\_num & \ccell{top1-5}{$1.53^{***}[13.53]$} & $1.06^{**\,\,\,}[\,\,\,0.20]$ & \ccell{top1-5}{$1.50^{***}[13.94]$} & \ccell{top1-5}{$1.50^{***}[\,\,\,8.01]$}  \\
   (4) & has\_comments & \ccell{top1-5}{$0.63^{***}[11.97]$} & \ccell{top1-5}{$0.52^{***}[25.39]$} & - & \ccell{top6-10}{$0.64^{***}[\,\,\,6.70]$}  \\
   (5) & core\_member & \ccell{top1-5}{$1.29^{***}[\,\,\,5.29]$} & $1.02^{\,\,\,\,\,\,\,\,\,}[\,\,\,0.05]$ & \ccell{top1-5}{$1.30^{***}[\,\,\,5.86]$} & \ccell{top6-10}{$1.32^{***}[\,\,\,3.58]$}  \\
   (6) & num\_commits & \ccell{top6-10}{$1.30^{***}[\,\,\,4.49]$} & \ccell{top1-5}{$1.35^{***}[\,\,\,6.67]$} & \ccell{top1-5}{$1.58^{***}[13.65]$} & \ccell{top1-5}{$1.56^{***}[\,\,\,7.43]$}  \\
   (7) & other\_comment & \ccell{top6-10}{$1.21^{***}[\,\,\,3.76]$} & \ccell{top1-5}{$1.27^{***}[\,\,\,6.47]$} & \ccell{top6-10}{$1.12^{***}[\,\,\,1.58]$} & \ccell{top6-10}{$1.24^{***}[\,\,\,2.88]$}  \\
   (8) & ci\_exists & \ccell{top6-10}{$1.16^{***}[\,\,\,1.47]$} & \ccell{top1-5}{$1.25^{***}[\,\,\,5.13]$} & \ccell{top6-10}{$1.11^{***}[\,\,\,0.93]$} & -  \\
   (9) & hash\_tag & \ccell{top6-10}{$1.12^{***}[\,\,\,1.36]$} & $1.06^{***}[\,\,\,0.51]$ & \ccell{top6-10}{$1.10^{***}[\,\,\,1.15]$} & \ccell{top6-10}{$1.13^{***}[\,\,\,1.04]$}  \\
   (10) & files\_added & \ccell{top6-10}{$0.91^{***}[\,\,\,0.74]$} & $0.96^{**\,\,\,}[\,\,\,0.18]$ & $0.91^{***}[\,\,\,0.61]$ & $0.90^{***}[\,\,\,0.48]$  \\
   (11) & prev\_pullreqs & $1.15^{***}[\,\,\,0.73]$ & $1.10^{***}[\,\,\,0.51]$ & \ccell{top6-10}{$1.16^{***}[\,\,\,0.99]$} & $1.15^{***}[\,\,\,0.43]$  \\
   (12) & commits\_on\_files\_touched & $1.09^{***}[\,\,\,0.59]$ & \ccell{top6-10}{$1.13^{***}[\,\,\,1.51]$} & $1.05^{***}[\,\,\,0.22]$ & $1.03^{***}[\,\,\,0.04]$  \\
   (13) & open\_pr\_num & $0.82^{***}[\,\,\,0.47]$ & $1.16^{***}[\,\,\,0.26]$ & $0.94^{***}[\,\,\,0.05]$ & $0.87^{***}[\,\,\,0.15]$  \\
   (14) & account\_creation\_days & $1.06^{***}[\,\,\,0.41]$ & \ccell{top6-10}{$1.16^{***}[\,\,\,2.52]$} & $1.06^{***}[\,\,\,0.41]$ & $1.09^{***}[\,\,\,0.53]$  \\
   (15) & first\_pr & $0.95^{***}[\,\,\,0.36]$ & $0.99^{\,\,\,\,\,\,\,\,\,}[\,\,\,0.01]$ & $0.96^{***}[\,\,\,0.27]$ & $0.96^{***}[\,\,\,0.16]$  \\
   (16) & test\_churn & $1.07^{***}[\,\,\,0.27]$ & \ccell{top6-10}{$1.10^{***}[\,\,\,0.59]$} & $1.11^{***}[\,\,\,0.59]$ & $1.12^{***}[\,\,\,0.42]$  \\
   (17) & files\_changed & $0.92^{***}[\,\,\,0.26]$ & $0.91^{***}[\,\,\,0.42]$ & $0.94^{***}[\,\,\,0.14]$ & $0.97^{***}[\,\,\,0.02]$  \\
   (18) & project\_age & $1.11^{***}[\,\,\,0.26]$ & $1.06^{\,\,\,\,\,\,\,\,\,}[\,\,\,0.08]$ & $1.08^{***}[\,\,\,0.19]$ & $1.21^{***}[\,\,\,0.53]$  \\
   (19) & reopen\_or\_not & $0.97^{***}[\,\,\,0.25]$ & $0.99^{\,\,\,\,\,\,\,\,\,}[\,\,\,0.05]$ & $0.98^{***}[\,\,\,0.12]$ & $0.98^{***}[\,\,\,0.08]$  \\
   (20) & contrib\_open & $1.06^{***}[\,\,\,0.24]$ & $1.05^{***}[\,\,\,0.27]$ & $1.05^{***}[\,\,\,0.20]$ & $1.07^{***}[\,\,\,0.20]$  \\
   (21) & stars & $0.86^{***}[\,\,\,0.22]$ & $0.88^{***}[\,\,\,0.22]$ & $0.79^{***}[\,\,\,0.69]$ & $0.89^{***}[\,\,\,0.10]$  \\
   (22) & inte\_open & $1.06^{***}[\,\,\,0.21]$ & $1.10^{***}[\,\,\,0.46]$ & $1.10^{***}[\,\,\,0.64]$ & $0.98^{***}[\,\,\,0.01]$  \\
   (23) & description\_length & $1.04^{***}[\,\,\,0.17]$ & $1.02^{\,\,\,\,\,\,\,\,\,}[\,\,\,0.04]$ & $1.01^{\,\,\,\,\,\,\,\,\,}[\,\,\,0.00]$ & $1.01^{***}[\,\,\,0.01]$  \\
   (24) & pushed\_delta & $1.04^{***}[\,\,\,0.15]$ & $1.06^{***}[\,\,\,0.39]$ & $1.04^{***}[\,\,\,0.22]$ & $1.04^{***}[\,\,\,0.12]$  \\
   (25) & followers & $1.04^{***}[\,\,\,0.12]$ & $0.92^{***}[\,\,\,0.52]$ & $1.02^{***}[\,\,\,0.04]$ & $1.03^{***}[\,\,\,0.03]$  \\
   (26) & contrib\_cons & $1.03^{***}[\,\,\,0.07]$ & $1.04^{**\,\,\,}[\,\,\,0.16]$ & $1.05^{***}[\,\,\,0.17]$ & $1.03^{***}[\,\,\,0.03]$  \\
   (27) & team\_size & $1.06^{***}[\,\,\,0.06]$ & $1.02^{\,\,\,\,\,\,\,\,\,}[\,\,\,0.00]$ & $1.06^{***}[\,\,\,0.07]$ & $1.07^{***}[\,\,\,0.06]$  \\
   (28) & contrib\_gender & $0.98^{***}[\,\,\,0.05]$ & $0.93^{***}[\,\,\,0.54]$ & $0.97^{***}[\,\,\,0.10]$ & $0.98^{***}[\,\,\,0.03]$  \\
   (29) & files\_deleted & $0.98^{***}[\,\,\,0.03]$ & $0.99^{\,\,\,\,\,\,\,\,\,}[\,\,\,0.02]$ & $0.96^{***}[\,\,\,0.18]$ & $0.96^{***}[\,\,\,0.10]$  \\
   (30) & pr\_succ\_rate & $0.98^{***}[\,\,\,0.03]$ & \ccell{top6-10}{$1.09^{***}[\,\,\,0.73]$} & $0.98^{***}[\,\,\,0.05]$ & $0.96^{***}[\,\,\,0.06]$  \\
   (31) & contrib\_agree & $0.98^{***}[\,\,\,0.02]$ & $0.99^{\,\,\,\,\,\,\,\,\,}[\,\,\,0.00]$ & $0.97^{***}[\,\,\,0.05]$ & $0.98^{***}[\,\,\,0.02]$  \\
   (32) & contrib\_extra & $0.99^{***}[\,\,\,0.02]$ & $0.94^{***}[\,\,\,0.29]$ & $0.97^{***}[\,\,\,0.07]$ & $0.97^{***}[\,\,\,0.04]$  \\
   (33) & contrib\_neur & $1.02^{***}[\,\,\,0.02]$ & $1.07^{***}[\,\,\,0.41]$ & $1.01^{**\,\,\,}[\,\,\,0.01]$ & $1.00^{\,\,\,\,\,\,\,\,\,}[\,\,\,0.00]$  \\
   (34) & inte\_neur & $1.02^{***}[\,\,\,0.02]$ & $1.00^{\,\,\,\,\,\,\,\,\,}[\,\,\,0.00]$ & $1.04^{***}[\,\,\,0.08]$ & $0.99^{\,\,\,\,\,\,\,\,\,}[\,\,\,0.00]$  \\
   (35) & num\_comments & $1.02^{***}[\,\,\,0.02]$ & $1.00^{\,\,\,\,\,\,\,\,\,}[\,\,\,0.00]$ & $0.91^{***}[\,\,\,0.88]$ & $0.97^{***}[\,\,\,0.04]$  \\
   (36) & comment\_conflict & $1.01^{***}[\,\,\,0.01]$ & $1.00^{\,\,\,\,\,\,\,\,\,}[\,\,\,0.00]$ & $1.02^{***}[\,\,\,0.05]$ & $1.01^{***}[\,\,\,0.01]$  \\
   (37) & friday\_effect & $1.01^{***}[\,\,\,0.01]$ & $1.01^{\,\,\,\,\,\,\,\,\,}[\,\,\,0.02]$ & $1.02^{***}[\,\,\,0.06]$ & $1.02^{***}[\,\,\,0.02]$  \\
   (38) & inte\_agree & $1.02^{***}[\,\,\,0.01]$ & $0.89^{***}[\,\,\,0.54]$ & $0.98^{***}[\,\,\,0.02]$ & $1.02^{*\,\,\,\,\,\,}[\,\,\,0.01]$  \\
   (39) & inte\_extra & $1.01^{***}[\,\,\,0.01]$ & $1.02^{\,\,\,\,\,\,\,\,\,}[\,\,\,0.01]$ & $1.01^{*\,\,\,\,\,\,}[\,\,\,0.01]$ & $1.06^{***}[\,\,\,0.10]$  \\
   (40) & open\_issue\_num & $1.03^{***}[\,\,\,0.01]$ & $1.08^{\,\,\,\,\,\,\,\,\,}[\,\,\,0.07]$ & $1.02^{\,\,\,\,\,\,\,\,\,}[\,\,\,0.00]$ & $1.03^{\,\,\,\,\,\,\,\,\,}[\,\,\,0.00]$  \\
   (41) & sloc & $1.02^{***}[\,\,\,0.01]$ & $0.97^{\,\,\,\,\,\,\,\,\,}[\,\,\,0.04]$ & $1.04^{***}[\,\,\,0.04]$ & $0.93^{***}[\,\,\,0.06]$  \\
   (42) & test\_inclusion & $1.02^{***}[\,\,\,0.01]$ & $1.00^{\,\,\,\,\,\,\,\,\,}[\,\,\,0.00]$ & $1.01^{*\,\,\,\,\,\,}[\,\,\,0.01]$ & $1.02^{***}[\,\,\,0.01]$  \\
   (43) & inte\_cons & $1.01^{\,\,\,\,\,\,\,\,\,}[\,\,\,0.00]$ & $1.04^{\,\,\,\,\,\,\,\,\,}[\,\,\,0.07]$ & $1.00^{\,\,\,\,\,\,\,\,\,}[\,\,\,0.00]$ & $0.99^{\,\,\,\,\,\,\,\,\,}[\,\,\,0.00]$  \\
   (44) & integrator\_availability & $1.00^{\,\,\,\,\,\,\,\,\,}[\,\,\,0.00]$ & $1.04^{**\,\,\,}[\,\,\,0.17]$ & $1.01^{**\,\,\,}[\,\,\,0.01]$ & $1.01^{\,\,\,\,\,\,\,\,\,}[\,\,\,0.00]$  \\
   (45) & src\_churn & $1.00^{\,\,\,\,\,\,\,\,\,}[\,\,\,0.00]$ & $1.00^{\,\,\,\,\,\,\,\,\,}[\,\,\,0.00]$ & $1.05^{***}[\,\,\,0.17]$ & $1.07^{***}[\,\,\,0.15]$  \\
   (46) & test\_lines\_per\_kloc & $1.01^{\,\,\,\,\,\,\,\,\,}[\,\,\,0.00]$ & $0.91^{***}[\,\,\,0.32]$ & $1.02^{***}[\,\,\,0.02]$ & $1.02^{*\,\,\,\,\,\,}[\,\,\,0.00]$  \\ [-0.5ex]

   \cmidrule(lr){1-6}\\[-3.3ex]

   (47) & agree\_diff & - & \ccell{top6-10}{$0.93^{***}[\,\,\,0.55]$} & - & -   \\
   (48) & cons\_diff & - & $0.98^{\,\,\,\,\,\,\,\,\,}[\,\,\,0.03]$ & - & -   \\
   (49) & contrib\_follow\_integrator & - & $1.01^{\,\,\,\,\,\,\,\,\,}[\,\,\,0.01]$ & - & -  \\
   (50) & extra\_diff & - & $0.99^{\,\,\,\,\,\,\,\,\,}[\,\,\,0.01]$ & - & -  \\
   (51) & neur\_diff & - & $0.98^{\,\,\,\,\,\,\,\,\,}[\,\,\,0.05]$ & - & -  \\
   (52) & open\_diff & - & $0.97^{*\,\,\,\,\,\,}[\,\,\,0.09]$ & - & -  \\
   (53) & same\_affiliation & - & $1.06^{***}[\,\,\,0.30]$ & - & -  \\
   (54) & same\_country & - & $1.02^{\,\,\,\,\,\,\,\,\,}[\,\,\,0.03]$ & - & -  \\
   (55) & perc\_pos\_emotion & - & - & \ccell{top6-10}{$1.18^{***}[\,\,\,3.14]$} & - \\
   (56) & perc\_neg\_emotion & - & - & $0.96^{***}[\,\,\,0.37]$ & - \\
   (57) & first\_response\_time & - & - & $1.01^{***}[\,\,\,0.02]$ & - \\
   (58) & ci\_failed\_perc & - & - & - & \ccell{top1-5}{$0.65^{***}[18.28]$}  \\  
   (59) & ci\_latency & - & - & - & \ccell{top6-10}{$1.11^{***}[\,\,\,0.67]$}  \\ 

   \hline \\[-2.5ex]
  & Observations & 1,765,730 & 91,874 & 839,505 & 954,386 \\
  & AUC\_train & 0.848 & 0.891 & 0.850 & 0.865 \\
  [-0.5ex]
  \bottomrule
  \end{tabular} 
  \end{table*} 


When the contributor and integrator are different users (\emph{same\_user=0}) (see column 4 in Table~\ref{result-special-case}), we found that three additional factor had a small effect in influencing pull request decision.
The only factor that made it to the top 10 factors is personality difference, and specifically differences in agreeableness (\emph{agree\_diff}). 
The two other factors are differences in openness to experience (\emph{open\_diff}), also indicating differences in personality, and same affiliation of the contributor and integrator (\emph{same\_affiliation}). 

When there exists at least one comment (\emph{has\_comments=1}) (see column 5 in Table~\ref{result-special-case}), what stands out is that positive emotion becomes relatively important with sizable effect ($>3\%$ variance). The change can be attributed to the phenomenon that positive reactions during the code review process can lead to contributor's active participation and increase the likelihood of pull request acceptance. While, for negative emotion, it's not important in pull request decision. A possible explanation for this is that different developers tend to act differently towards negative emotion. Therefore, negative emotion during discussion is difficult to effectively distinguish the final decision.
To verify our observation, we built models for pull requests that have at least one comment from contributor (\emph{contrib\_comment=1}) or at least one comment from integrator (\emph{inte\_comment=1})~\cite{Zhang-tse-report}.
We found that both \emph{perc\_contrib\_pos\_emo} and \emph{perc\_inte\_pos\_emo} explain more than 3\% of the variance and much higher than negative emotion.

When pull requests use CI tools (\emph{ci\_exists=1}) (see column 6 in Table~\ref{result-special-case}), factor \emph{ci\_failed\_perc} stands out, explaining 18\% of the variance. It implies that the build status of CI tools is important for review decision, especially the percentage of build failure.

\begin{tcolorbox}
 Pull request decision is mostly explained by a few factors (5 to 10 factors) such that developer and pull request characteristics are more important than project characteristics.
 
The relation between contributor and integrator (\emph{same\_user}) is the most important factor influencing pull request decision. 

In special cases, when a pull request has comments, comment's positive emotion is linked to pull request acceptance.
Likewise, when pull requests use CI tools, the percentage of failed CI builds become important for pull request decision.
\end{tcolorbox}

\subsection{RQ2: Do factors influencing pull request decisions change with context?}
\label{Result_RQ3}

\begin{table*}[!h] \centering
  \setlength\tabcolsep{2pt}
  \scriptsize
  \caption{presents part of the results in different contexts. \\ We keep outstanding results here, the full table can be seen in Appendix~\ref{appendix-result-different-contexts}. \\
  Gray color marks the factors that have more than 5\% difference of explained variance in different contexts.\\
  Value before bracket means the odds ratio, value in bracket means the percentage of explained variance, - means the factor is not included in the model} 
  \label{result-different-contexts-part} 
\begin{tabular}{ clcccccccccccc} 
\\[-1.8ex]
\toprule
\\[-3ex] 
 & (1) & (2) & (3) & (4) & (5) & (6) & (7) & (8) & (9) & (10) & (11) & (12) & (13) \\
 & \multicolumn{13}{c}{\textit{Dependent variable: merged\_or\_not=1}} \\ 
\cline{2-14} 
\\[-1.8ex] 
 & & \multicolumn{2}{c}{same user or not} & \multicolumn{2}{c}{has comments or not} & \multicolumn{2}{c}{ci exists or not} & \multicolumn{3}{c}{different team sizes} & \multicolumn{3}{c}{different periods}\\
 & & yes & no & yes & no & yes & no & small & mid & large & before 2016.6 & 2016.6-2018.6 & after 2018.6 \\
 \cmidrule(lr){3-4} \cmidrule(lr){5-6} \cmidrule(lr){7-8} \cmidrule(lr){9-11} \cmidrule(lr){12-14}\\[-1.8ex] 
 & (Intercept) & 10.4 & 34.4 & 13.1 & 42.4 & 20.4 & 13.3 & 24.9 & 20.7 & 15.9 & 6.9 & 16.6 & 7.1 \\ 
 
 (1) & prior\_review\_num & \ccell{cellgray}{$2.86[31]$} & \ccell{cellgray}{$0.98[\,\,\,0]$} & \ccell{cellgray}{$1.51[14]$} & \ccell{cellgray}{$1.91[22]$} & $1.53[14]$ & $1.53[12]$ & \ccell{cellgray}{$1.59[11]$} & \ccell{cellgray}{$1.41[\,\,\,9]$} & \ccell{cellgray}{$1.57[19]$} & \ccell{cellgray}{$1.30[\,\,\,6]$} & \ccell{cellgray}{$1.63[14]$} & \ccell{cellgray}{$1.72[17]$} \\

(2) & lifetime\_minutes & \ccell{cellgray}{$0.66[19]$} & \ccell{cellgray}{$0.52[44]$} & \ccell{cellgray}{$0.61[30]$} & \ccell{cellgray}{$0.70[13]$} & $0.60[22]$ & $0.61[21]$ & $0.54[24]$ & $0.61[20]$ & $0.67[17]$ & \ccell{cellgray}{$0.65[20]$} & \ccell{cellgray}{$0.57[21]$} & \ccell{cellgray}{$0.62[13]$} \\

 (3) & core\_member & \ccell{cellgray}{$1.26[\,\,\,9]$} & \ccell{cellgray}{$1.13[\,\,\,1]$} & $1.29[\,\,\,6]$ & $1.33[\,\,\,6]$ & $1.30[\,\,\,5]$ & $1.26[\,\,\,5]$ & $1.42[\,\,\,6]$ & $1.28[\,\,\,5]$ & $1.19[\,\,\,3]$ & $1.27[\,\,\,6]$ & $1.34[\,\,\,5]$ & $1.29[\,\,\,3]$ \\
 
 (4) & num\_commits & \ccell{cellgray}{$1.23[\,\,\,4]$} & \ccell{cellgray}{$1.46[10]$} & \ccell{cellgray}{$1.49[11]$} & \ccell{cellgray}{$0.98[\,\,\,0]$} & $1.32[\,\,\,5]$ & $1.25[\,\,\,4]$ & $1.36[\,\,\,5]$ & $1.31[\,\,\,5]$ & $1.26[\,\,\,4]$ & $1.18[\,\,\,2]$ & $1.32[\,\,\,4]$ & $1.36[\,\,\,5]$ \\

 (5) & commits\_on\_files\_touched & $1.06[\,\,\,0]$ & $1.11[\,\,\,1]$ & $1.05[\,\,\,0]$ & $1.18[\,\,\,2]$ & $1.06[\,\,\,0]$ & $1.13[\,\,\,1]$ & $1.10[\,\,\,0]$ & $1.12[\,\,\,1]$ & $1.05[\,\,\,0]$ & \ccell{cellgray}{$1.30[\,\,\,7]$} & \ccell{cellgray}{$0.99[\,\,\,0]$} & \ccell{cellgray}{$0.99[\,\,\,0]$} \\


 (6) & has\_comments & \ccell{cellgray}{$0.68[10]$} & \ccell{cellgray}{$0.50[27]$} & - & - & \ccell{cellgray}{$0.65[10]$} & \ccell{cellgray}{$0.52[25]$} & $0.57[13]$ & $0.64[10]$ & $0.66[12]$ & \ccell{cellgray}{$0.63[15]$} & \ccell{cellgray}{$0.62[10]$} & \ccell{cellgray}{$0.55[14]$} \\

 (7) & same\_user & - & - & \ccell{cellgray}{$0.56[29]$} & \ccell{cellgray}{$0.42[42]$} & \ccell{cellgray}{$0.51[33]$} & \ccell{cellgray}{$0.59[23]$} & \ccell{cellgray}{$0.49[24]$} & \ccell{cellgray}{$0.49[36]$} & \ccell{cellgray}{$0.55[33]$} & $0.57[31]$ & $0.46[33]$ & $0.46[29]$ \\

 & \vdots & \vdots & \vdots & \vdots & \vdots & \vdots & \vdots & \vdots & \vdots & \vdots & \vdots & \vdots & \vdots \\
  
 \hline \\[-1.8ex]
\multicolumn{2}{l}{Observations} & 950,985 & 1,010,937 & 1,152,714 & 809,208 & 1,611,277 & 350,645 & 601,460 & 703,396 & 701,900 & 512,707 & 585,401 & 274,121 \\
\multicolumn{2}{l}{AUC\_train} & 0.862 & 0.874 & 0.837 & 0.872 & 0.843 & 0.884 & 0.877 & 0.843 & 0.837 & 0.850 & 0.867 & 0.879 \\
\bottomrule
\\[-1.8ex]
\end{tabular}
\end{table*}

\subsubsection{Developer characteristic}
\label{developer-characteristic}

Table~\ref{result-different-contexts-part} shows that in comparison to the pull requests submitted and integrated by the same user, when the contributor and integrator are not the same person, the variance explained by the experience of integrator (\emph{prior\_review\_num}) decreases from 31\% (row 1, column 2 - same user: yes) to 0\% (row 1, column 3 - same user: no). 
This implies that integrator's experience has limited role when making decision on others' contributions.
However, the factor becomes very important for their own contributions.
One way to explain the observation can be that external contributors, without review experience, generally do not have the right to merge the code.
Experienced integrators, in contrast, are familiar with the management process, know when to merge a pull request, and can merge a pull request.
This way, difference in permission linked to integrator's experience can influence pull request decision.

For the lifetime of pull requests (\emph{lifetime\_minutes}), the percentage of explained variance increases from 19\% (row 2, column 2 - same user: yes) to 44\% (row 2, column 3 - same user: no). 
A possible explanation for this observation is that when there is no response from contributor for a long time, it is more likely to be closed by reviewer. However, when the pull request is reviewed by the contributor themself, they know exactly what is going on and the decision making will not be influenced as much by the indicator lifetime of a pull request.
Likewise, for the number of commits (\emph{num\_commits}), the percentage of explained variance increases from 4\% (row 4, column 2 - same user: yes) to 10\% (row 4, column 3 - same user: no). It's likely that during the interaction, the integrator will ask the contributor to modify the contribution, increasing the number of commits, and then make decision according to the changes.

When comments are present (\emph{has\_comments}), the explained variance increases when a pull request is integrated by another person in comparison to self from 10\% (row 6, column 2 - same user: yes) to 27\% (row 6, column 3 - same user: no). The result can be explained by the fact that when integrating pull requests submitted by others, it's common to understand the contribution by communicating with the contributor.

When a contributor is a core member (\emph{core\_member}), the explained variance decreases from 9\% (row 3, column 2 - same user: yes) to 1\% (row 3, column 3 - same user: no). There are two likely causes for the observation. First, if a contributor is a core member, they can merge their pull requests. Others can only close their own contributions. Second, when reviewing others' pull requests, reviewers are less likely to take a decision simply because the contributor is a core member.

\begin{tcolorbox}
    Whether the contributor and integrator is the same person or not influences the pull request decision the most.
    
If the contributor and integrator is the same, pull request decision depends on the contributor's relationship to the target project (\emph{prior\_review\_num} and \emph{core\_member}).

When the contributor and integrator are different, pull request decision depends on the interaction between contributor and integrator (\emph{has\_comments}, \emph{lifetime\_minutes}) and the intermediate result during the process (\emph{num\_commits}).
\end{tcolorbox}

\subsubsection{Pull request characteristic}
\label{pull-request-characteristic}

When a pull request does not have comment(s), the percentage of explained variance of factor \emph{same\_user} increases from 29\% (row 7, column 4 - has comments: yes) to 42\% (row 7, column 5 - has comments: no).
One possible explanation for the observation is that pull request decision depends on reviewer's comments. However, when there is no comment, it is likely that either contributor has found a problem in the contribution and closes it themselves or the pull request is well understood and easy to merge by others.

For the experience of integrator (\emph{prior\_review\_num}), the explained variance increases from 14\% (row 1, column 4 - has comments: yes) to 22\% (row 1, column 5 - has comments: no). It is likely that when there is no comment, there are cases in which developers close or merge their own pull requests. 
In comparison to core members, external developers do not have right to merge.
The restricted permission linked to integrator's review experience can potentially influence the pull request decision.

For the lifetime of a pull request (\emph{lifetime\_minutes}) and the number of commits included in pull request (\emph{num\_commits}), when there exists comment(s), potentially the integrator tend to make decision based on contributor's response speed and how they modify the contribution according to integrator's suggestions.
This can be a reason why there exists a higher percentage of variance in situation when there exists comment.

\begin{tcolorbox}
    When there is no communication between contributor and reviewer, factors indicating the affiliation of a contributor to the project - 
    whether the contributor and reviewer are same (\emph{same\_user}) and review experience (\emph{prior\_review\_num}), are important in influencing pull request decision.

When there is communication between contributor and reviewer, factors representing the activeness of the interaction (\emph{lifetime\_minutes}, \emph{num\_commits}) have a bigger influence on pull request decision.
\end{tcolorbox}

\subsubsection{Project characteristic}
\label{project-characteristic}
As the team size increases, the variance explained by the experience of integrator (\emph{prior\_review\_num}) initially decreases from 11\% (row 1, column 8 - team size: small) to 9\% (row 1, column 9 - team size: mid) and then increases from 9\% (row 1, column 9 - team size: mid) to 19\% (row 1, column 10 - team size: large).

When considering whether pull requests are submitted and integrated by the same user (\emph{same\_user}), the change trend is the opposite, increasing from 24\% (row 7, column 8 - team size: small) to 36\% (row 7, column 9 - team size: mid), and then decreasing from 36\% (row 7, column 9 - team size: mid) to 33\% (row 7, column 10 - team size: large).

The two types of change indicate that for pull requests targeting different sized teams, the importance of factor \emph{prior\_review\_num} and \emph{same\_user} changes non-linearly.
We do not have any reason to explain the observation.

\begin{tcolorbox}
    As team size increases, integrator's experience (\emph{prior\_review\_num}) and whether submitter and integrator is the same (\emph{same\_user}) have a V-shaped and inverted V-shaped relations to the pull request decision respectively.
\end{tcolorbox}

\subsubsection{Supporting tools}
\label{supporting-tools}

When not using CI tools, the percentage of variance explained by comments (\emph{has\_comments}) is 25\% (row 6, column 7 - ci exists: no), which is higher than that of pull requests using CI tools (10\%) (row 6, column 6 - ci exists: yes).
The result can be explained by the fact that when there is no CI tools, contributors can only get feedback from reviewers. Therefore, whether there exists comment or not matters a lot in pull request decision. 
When using CI tools, contributors can firstly get response from CI outcome, which can help with decision.

When pull requests use CI tools, the contributor can directly get the build outcome and decide whether to close or merge a pull request. However, when there is no CI tools, it relies more on the manual review. Therefore, the explained variance of whether contributor and integrator are the same user (\emph{same\_user}) decreases from 33\% (row 7, column 6 - ci exists: yes) to 23\% (row 7, column 7 - ci exists: no).

\begin{tcolorbox}
    CI tool automates the code review process, replacing some code inspection work, accelerates the review process, and assists decision making.
\end{tcolorbox}

\subsubsection{Project Evolution}
\label{project-development}

Before 2016.6, the experience of integrator (\emph{prior\_review\_num}) explains just 6\% (row 1, column 11 - period: before 2016.6) of the variance which increases to 17\% after 2018.6 (row 1, column 13 - period: after 2018.6).
One possible explanation is that when projects become mature, relatively more experienced integrators now can easily guide contributors and help them modify pull requests to meet project's standards. However, in early stage, there is no big difference in review experience among integrators. This can be the reason why factor \emph{prior\_review\_num} does not play a decisive role in pull request decision in early time periods.  

For the area hotness of contributions (\emph{commits\_on\_files\_touched}), before 2016.6, it has moderate effect on the decision making of pull request, which explains 7\% of the variance (row 5, column 11 - period: before 2016.6), and increases the odds of acceptance by 30\% per unit. However, as projects become mature, the variance explained decreases to 0\% (row 5, column 13 - period: after 2018.6).
For the three periods, we also calculated the mean value of \emph{commits\_on\_files\_touched} (before 2016.6: 40, 2016.6-2018.6: 33, after 2018.6: 28), which shows that the contributions in the early stage of the project are more concentrated.
In other words, as projects become larger and more mature, contributions are more widely distributed in the project, and the area hotness of pull request can hardly contribute to the merge of pull request for mature projects.

For the lifetime of pull request (\emph{lifetime\_minutes}), the explained variance decreases from 20\% (row 2, column 11 - period: before 2016.6) to 13\% (row 2, column 13 - period: after 2018.6).
We calculate the mean value of workload (\emph{open\_pr\_num}) (before 2016.6: 58, 2016.6-2018.6: 112, after 2018.6: 174), it increases as projects become mature.
Perhaps, the increase of workload brings an overall improvement in pull request lifetime which likely decreases the importance of lifetime in pull request decision.


\begin{tcolorbox}
  As project evolves, integrator's experience (\emph{prior\_review\_num}) becomes more and more important for pull request decision, while the area hotness of contribution (\emph{commits\_on\_files\_touched}) no longer influences decision.
  
 Compared to the early stages of project evolution, the influence of pull request lifetime (\emph{lifetime\_minutes}) on pull request decision decreases. 
\end{tcolorbox}

\section{Discussion}
\label{discussion}

\subsection{Pull request decision explained}




Our study shows that there is no one answer to this question. 
Instead, there is a generic answer and a specific answer for the context it represents, given the dependencies among factors.
Generally, whether a pull request is submitted and integrated by the same person, its lifetime, experience of integrator, presence of comment(s), and coreness of the contributor play decisive roles in pull request decision.
When there exists \emph{comment in pull requests}, the positive emotion for communication influences pull request decision.
While when pull requests use \emph{CI tool(s)}, the percentage of build failure influences the decision.

Interestingly, the influence of the factors changes with context:

\emph{Developer characteristic (same user or not)}: Comparing to pull requests integrated by different persons, when pull requests are submitted and integrated by the same person, the importance of integrator's experience and contributor's coreness increase for pull request decision, while the importance of pull request lifetime and the included number of commits decrease.

\emph{Pull request characteristic (has comment or not)}: When pull requests have comments, lifetime and the number of commits included are more important comparing to pull requests without any comment. Contrarily, the importance of integrator's experience and whether contributor and integrator are the same person change are less important when there exists comment.

\emph{Tool (CI exists or not)}: The use of CI tools decreases the importance of comment existence, but the importance of whether contributor and integrator are the same person increases for pull request decision.

\emph{Project characteristics (different team size)}: The importance of integrator's experience and whether contributor and integrator are the same person for pull request decision changes non-linearly for different sized teams.

\emph{Project evolution (different periods)}: The importance of integrator's experience on pull request decision increases as project evolves, while the importance of area hotness and lifetime of contribution decreases.

\subsection{Relations to literature}

Referring to the literature (summarized in Table~\ref{factor_table}), relatively speaking project-related factors are less discussed than pull request and developer-related factors.
To this, our study adds that not only few project characteristics are explored in literature, they are relatively less important (explains 2\% variance) in comparison to developer (explains 52\% variance) and pull request characteristics (explains 46\% variance).
Our study further gives an evidence to the observation  that human factors  are as important or more important  than technical factors~\cite{hoc2014psychology}.

When comparing the findings of these studies to each other and our study, we found that in most of the cases, the results are consistent. Only four factors had opposite findings regarding the direction of influence, \emph{i.e.} \emph{files\_changed}, \emph{project\_age}, \emph{team\_size} and \emph{num\_commits}. One potential explanation that emerged from our study is that all these factors are relatively less important for pull request decision which can potentially explain the differences in finding. 
Alternatively, this can simply be due to the difference in dataset.
Interestingly, many factors that are widely studied across related works, \emph{e.g.} \emph{core\_member} and \emph{src\_churn}, indicating that these factors are likely to influence the decision, are not so important for pull request decision.

\subsection{Implications}

Our findings have implications for research, practice, and education.
First, when doing research on pull request decision, one can find usable findings from our paper for both a general overview and specific contexts.
Second, future researchers can use and extend our large and rich dataset\footnote{https://zenodo.org/record/4837134\#.YLEWyY3isdW} for deeper investigations and use the scripts to replicate the results.\footnote{https://github.com/zhangxunhui/TSE\_pull-based-development}
Third, our study offers recommendations for contributors to improve pull request decision. For example, our study shows that contributors should focus more on controlling the processing time and speed up interaction with reviewers for improved chances of pull request acceptance.
Lastly, now that pull-based development is a standard collaborative development process, it is necessary to understand and educate how pull request decisions are made. 


\section{Threats to Validity}
\label{threats}
Our work builds on a decade of research on pull-based development, extracting features relevant for decision making of pull request. 
This way we not only stand on the shoulders of giants, and hence benefiting from it but also inherit limitations of the features they present.
In addition, we have the following limitations.

\begin{enumerate}
    \item There are different ways to calculate the importance of factors in logistic regression model, \emph{e.g.} the percentage of variance explained by each factor~\cite{perc-variance}, which is similar to the percentage of total variance explained by least square regression~\cite{cohen_applied}, standardized coefficient~\cite{relative_importance}, the change of logistic pseudo partial correlation~\cite{variable-importance-so}, etc. This is a research field in itself, and relates to the choice of algorithm~\cite{variable-importance-so-algo1, variable-importance-so-algo2}.
In order to compare the importance of factors in different models, in this paper we choose the percentage of explained variance to represent factors' importance.
    
    \item For logistic regression models, comparing the variance explained by the same factor in different models is not accurate, as they use different training set. However, when building models with the same set of predictors, big changes of explained variance can be used to describe the change in factor's importance to some extent. Therefore, in our study, we just consider the factors that changes dramatically in different contexts.
    
    \item Even though we have collected a large dataset, we still cannot completely avoid the data bias when building up models in different contexts. Although the diversity of our data can avoid this problem to some extent.
    
    \item During the data preprocessing, we remove factor \emph{bug\_fix} due to many missing value. We are not sure how this factor will affect the pull request decision making.
    
    \item We are not sure how our result apply to other social coding platforms (other than GitHub). 
 One major reason for the differences can be the factors influencing pull request decision on different platforms. 
    
    \item We considered factors that can be mined from archival data. We excluded factors, \emph{e.g.,} the eye-track related factors\cite{Ford_eye} that are difficult to quantify in a scalable manner. This also include factors that only focuses on specific scenarios, \emph{e.g.,} factors related to Microsoft\cite{Maddila_latency}, NPM ecosystem only\cite{Dey_npm}, etc. We are not sure how these factors perform along with our collected factors.
    
\end{enumerate}



\section{Related Work}
\label{background}

The related work of this paper is mainly divided into three parts.
The first subsection introduces factors influencing pull request decision.
Secondly, we introduce papers that tried to integrate related factors and explain the relative importance of factors influencing pull request decision.
Thirdly, we introduce other studies introduce scientific research methods based on big data.

\subsection{Factors influencing pull request decision}
Factors influencing pull request decision can be divided into three categories, namely developer characteristics, project characteristics and pull request characteristics.

\subsubsection{Developer Characteristics}
The developer characteristics are related to the contributor and integrator. Some of the factors are human related while others represent interactions between two contributors or a contributor and a project.
This category include some \texttt{basic information} of developers, including the \emph{gender}~\cite{Terrell_gender}, \emph{country} information~\cite{Rastogi_relationshipESEM}, and \emph{affiliation} of developers~\cite{Kononenko_shopify, Baysal_Nontech}.
Some studies focus on the \texttt{personal features}, including \emph{personality} and \emph{emotion} of developers~\cite{Iyer_tse,Iyer_master}.
Others studied the \texttt{relationship} between developer and the target project, including the \emph{experience} of developers, which is conceptualized as the count of previous pull requests, accepted commit count~\cite{Jiang_linux}, days since account creation~\cite{Rahman_insight}, whether it's the first pull request of contributor~\cite{Soares_rejection, Soares_acceptance}, as well as the prior reviews of integrator~\cite{Baysal_Nontech},
the \emph{coreness} of contributor~\cite{Tsay_influence, Yu_determinants, Soares_acceptance, Pinto_who, Baysal_firefox, Bosu_reputation},
the \emph{social distance}~\cite{Tsay_influence} and \emph{social strength}~\cite{Yu_determinants} of contributor to the integrator,
and the \emph{response time} of integrator to pull request~\cite{Yu_determinants}.

\subsubsection{Project characteristics}
Studies on project characteristics mainly talked about the \texttt{basic information} of target projects when submitting pull requests, which can be summarized into the following aspects:
\emph{programming language}~\cite{Padhye_external, Soares_acceptance, Rahman_insight},
\emph{popularity} of project, measured as watcher count~\cite{Gousios_Dataset}, star count~\cite{Gousios_Dataset}, and fork count~\cite{Khadke_predict, Rahman_insight}, 
\emph{age} of project~\cite{Tsay_influence, Yu_determinants},
\emph{workload} measured as the number of open pull requests~\cite{Baysal_Nontech, Yu_determinants},
\emph{activeness} measured as the time interval in seconds between the opening time of two latest pull requests~\cite{Khadke_predict},
and \emph{openness} measured as the count of open issues~\cite{Khadke_predict}.

\subsubsection{Pull request characteristics}
Related works focus on the \texttt{basic information} of pull request, which includes the \emph{size of change} measured at file level, commit level, and code level~\cite{Yu_determinants}, 
the \emph{complexity of a pull request} measured as the length of description~\cite{Yu_determinants},
the \emph{nature of pull request} as bug fix~\cite{Padhye_external, Jiang_linux},
the \emph{test inclusion} of pull requests~\cite{Yu_determinants, Tsay_influence, Pinto_who},
the \emph{hotness} or relevance of a PR ~\cite{Yu_determinants, Gousios_exploratory, Tsay_influence, Soares_rejection, Kononenko_shopify, Jiang_linux}.
Also, some studies focus on the \texttt{process information} of pull request generated during code review process, including
the \emph{reference} of a contributor, issue or pull request~\cite{Yu_determinants, Calefato_trust_2017},
the \emph{conflict} of a pull request~\cite{Gousios_exploratory},
the \emph{complexity of discussion}~\cite{Gousios_Dataset},
the \emph{emotion in discussion}~\cite{Iyer_master},
the \emph{CI tool usage} during the review process~\cite{Yu_determinants, Zampetti_ci, Bogdan_CI, Gousios_integrator, Tao_Empirical}. 

\subsection{Attempts at explaining pull request decision}
Few studies tried to integrate factors related to pull request decision and explored their relative importance in predicting the outcome.
Gousios et al.~\cite{Gousios_exploratory} firstly collected a set of factors and did a preliminary exploration of relative importance based on random forest method.
But it was in the early stage of this study area. 
Tsay et al.~\cite{Tsay_influence} used explanatory method to explore the importance of social and technical factors. However, similar to Gousios et al.'s work~\cite{Gousios_exploratory}, their work also acted as a groundbreaking research, and many studies emerged after that.
Since then a few follow-ups come into being, \emph{e.g.} the personality related factors~\cite{Iyer_tse}, geographical location~\cite{Rastogi_relationshipESEM}, CI related factors~\cite{Yu_determinants}, etc.
In 2020, Dey et al.~\cite{Dey_npm} collected 50 factors of 483,988 pull requests based on 4218 projects. They also used random forest method to find out the important factors in predicting the decision.
However, they just focused on NPM community and gathered factors without conducting systematic literature review process.
As a result, factors related to CI, personality, emotion, geographical, etc were missing.
Further, to the best of our knowledge no study synthesized the existing body of knowledge to empirically explain pull request decision.

\subsection{Big data based scientific research methods}
Big data has provided many research opportunities for conducting research, and there are mainly two research methods, \emph{i.e.} data-driven and theory-driven methods.
Maass et al.~\cite{Maass-big-data} discussed the difference between these two methods, and found that for data-driven method, it firstly focus on the data, then extract patterns and forms into theory. However, for theory-driven method, it firstly comes up with the theory and uses data to prove.
Therefore, our study is data-driven which finds patterns in different subsets of data and form into theory in the end.

For the process of data-driven study, Kar et al.~\cite{Kar-data-driven} described that there are mainly 6 steps for building up theory in data-driven research, \emph{i.e.} data acquisition, data conversion, data analysis, factor identification, theory development and model validation.

There are many studies in different research areas using data-driven research method.
For example, Greenwood et al.~\cite{Greenwood-data-driven} studied the influence of race, gender, and socioeconomic status on the incidence rate of human immunodeficiency virus (HIV) infection using the data from 12 million patients.
Likewise, previous studies~\cite{Gousios_exploratory,Yu_determinants,Tsay_influence,Rastogi_relationshipESEM} about pull request decision all used data-driven method.

However, for the data acquisition part , previous studies only focused on one specific type of factor or several self-defined factors.
Without including all the related factors, one can hardly form an overall grasp of the influence of all factors.
Therefore, we conduct a systematic literature review for this study.
According to Kitchenham et al.~\cite{Kitchenham-ebse}, systematic literature review is an important part in evidence based software engineering (EBSE) as it can aggregate all existing evidence and provide guidelines for researchers.

\section{Conclusion}
\label{conclusion}
This study synthesizes the existing body of knowledge to empirically explain pull request decision.
Our mixed-effect logistic regression models built on a large and diverse GitHub projects' data shows that a handful of factors (5 to 10) explain pull request decision the most.
The most important factor influencing pull request decision is whether contributor and integrator is the same user, explaining more than 30\% of the variance. 
Surprisingly, this factor did not surface in any of the prior works and is also a contribution of this study.
In addition, positive emotions during discussion and CI build results become relatively more important when a pull request has comments and uses CI tools respectively. 
Further, we noticed that using CI tools replaces the function of comments indicating changes in the influence of factors. 
We believe that this study has now synthesized an explanation for pull request decision empirically that is useful for research, practice, and education.


\section*{Acknowledgment}
This work is supported by 
Science and Technology Innovation 2030
of China (Grand No.2018AAA0102304).
Thank you Rahul N. Iyer, Frenk van Mil, Celal Karakoc, Leroy Velzel, Daan Groenewegen, and Sarah de Wolf for your help in implementation.

\bibliographystyle{./bibliography/IEEEtran}
\bibliography{./bibliography/main}

\begin{IEEEbiography}[{\includegraphics[width=1in,height=1.25in,clip,keepaspectratio]{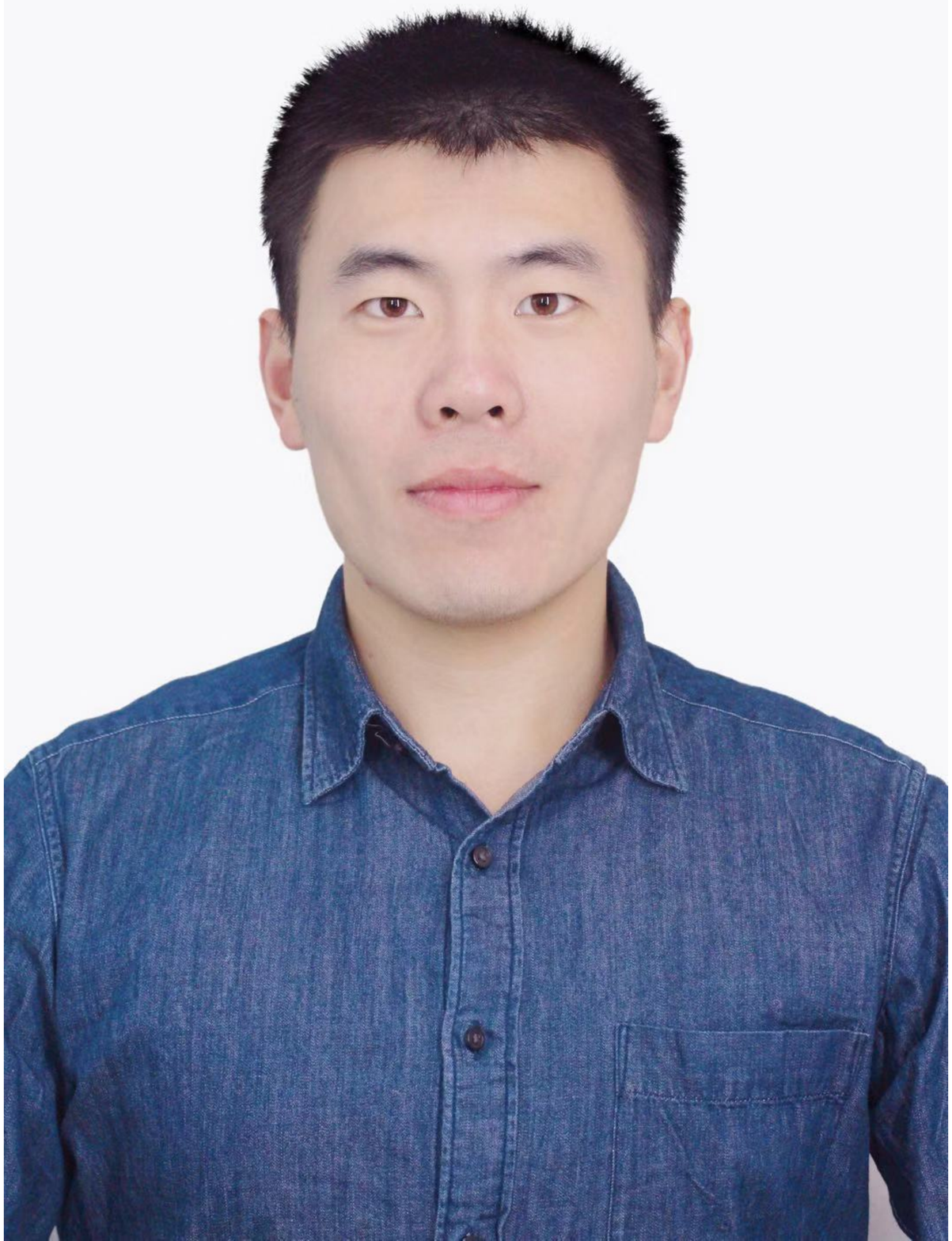}}]{Xunhui Zhang}
received his BS in Computer Science from Sichuan University in 2015. He received his MS in Software Engineering from National University of Defense Technology in 2017. He is now a PhD candidate in Software Engineering, National University of Defense Technology. His work interests include open source software engineering, data mining, recommendation system, cross community analysis and code clone.
\end{IEEEbiography}

\begin{IEEEbiography}[{\includegraphics[width=1in,height=1.25in,clip,keepaspectratio]{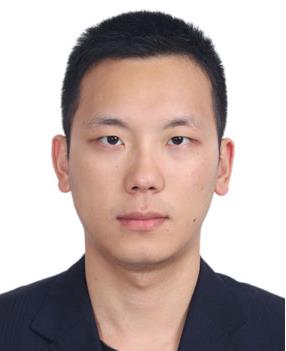}}]{Yue Yu}
  is an associate professor in the College of Computer at National University of Defense Technology (NUDT). 
He received his Ph.D. degree in Computer Science from NUDT in 2016. 
He has won Outstanding Ph.D. Thesis Award from Hunan Province.
His research findings have been published on
ICSE, FSE, ASE, TSE, MSR, IST, ICSME, ICDM and ESEM.
His current research interests include software engineering, data mining and computer-supported cooperative work.
\end{IEEEbiography}

\begin{IEEEbiography}[{\includegraphics[width=1in,height=1.25in,clip,keepaspectratio]{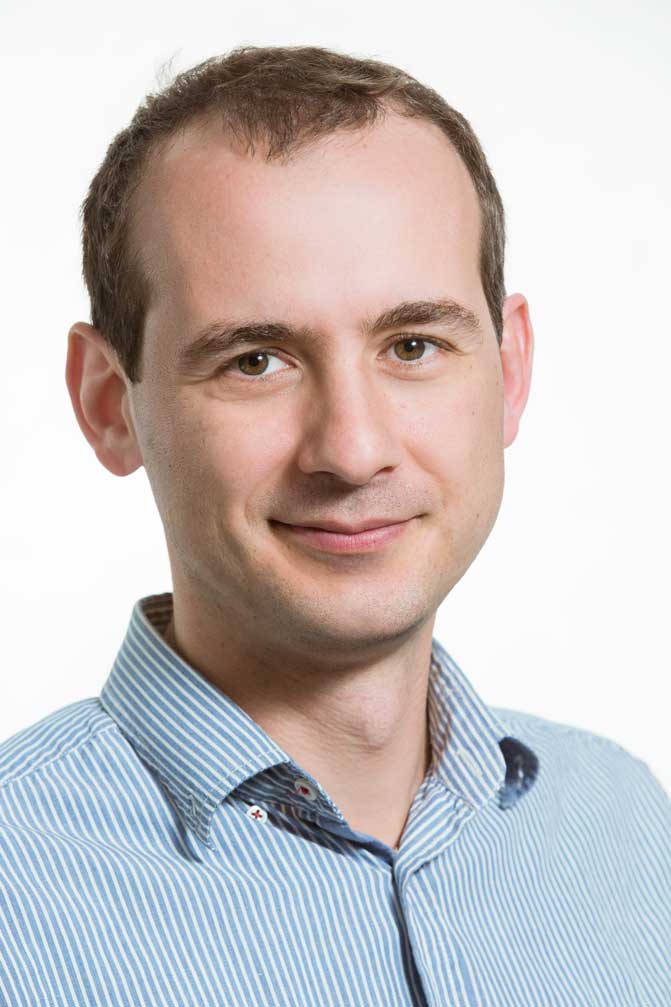}}]{Georgios Gousios}
  is a research scientist at Facebook and an associate professor at the Delft University of Technology (on leave). His work is on applying techniques from the domains of static analysis, machine learning and software analytics to improve developer productivity and operational efficiency.
\end{IEEEbiography}

\begin{IEEEbiography}[{\includegraphics[width=1in,height=1.25in,clip,keepaspectratio]{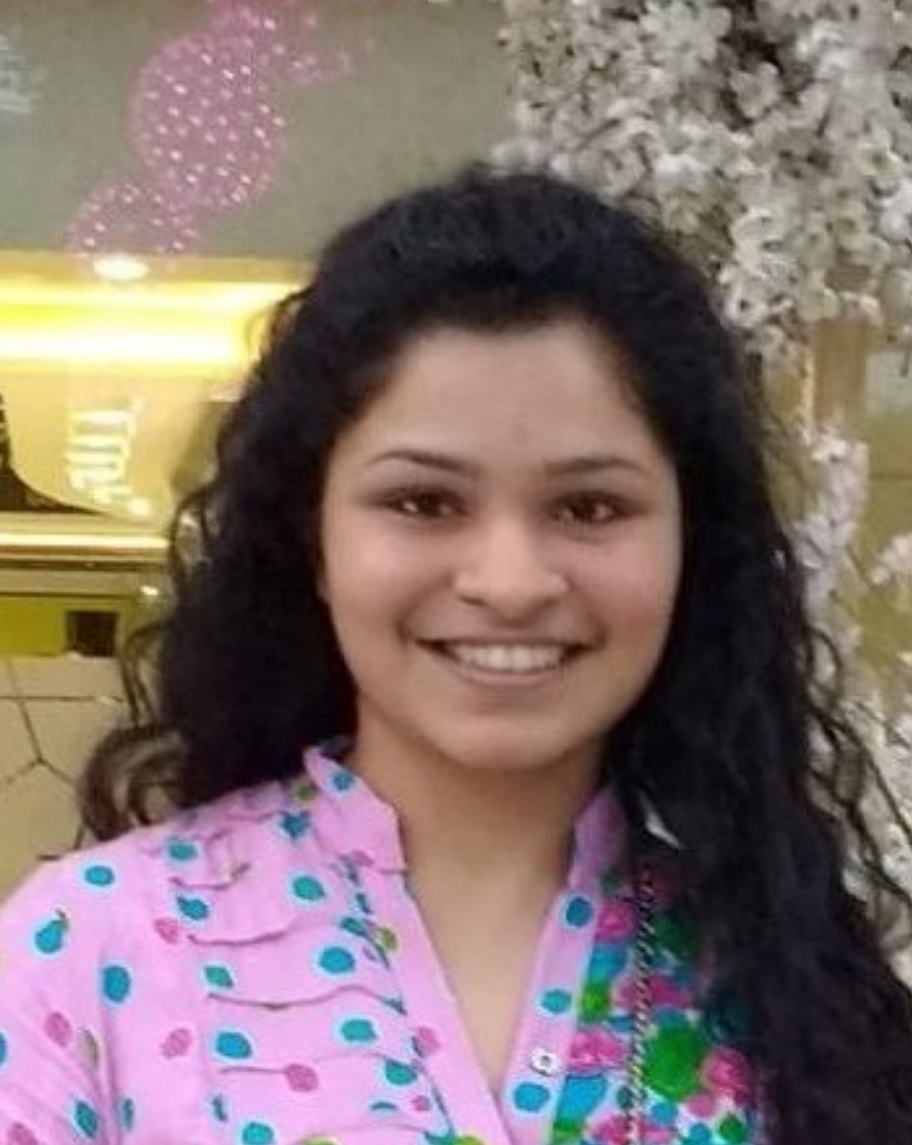}}]{Ayushi Rastogi}
  is an Assistant Professor in the Faculty of Science and Engineering at the University of Groningen, the Netherlands. Her research interests include software analytics, empirical software engineering, and mining software repositories. She studies human and social aspects of software engineering for improving developer productivity and promoting diversity and inclusion.
\end{IEEEbiography}

\clearpage
\onecolumn
\begin{landscape}
\setlength{\tabcolsep}{0.1em}
\footnotesize
\renewcommand{\arraystretch}{1.37}
\begin{longtable}{l >{\centering}m{0.75cm} >{\centering}m{0.75cm} >{\centering}m{0.75cm} >{\centering}m{0.75cm} >{\centering}m{0.75cm} >{\centering}m{0.75cm} >{\centering}m{0.75cm} >{\centering}m{0.75cm} >{\centering}m{0.75cm} >{\centering}m{0.75cm} >{\centering}m{0.75cm} >{\centering}m{0.75cm} >{\centering}m{0.75cm} >{\centering}m{0.75cm} >{\centering}m{0.75cm} >{\centering}m{0.75cm} >{\centering}m{0.75cm} >{\centering}m{0.75cm} >{\centering}m{0.75cm} >{\centering}m{0.75cm} >{\centering}m{0.75cm} >{\centering}m{0.75cm} >{\centering}m{0.75cm} >{\centering}m{0.75cm} >{\centering}m{0.75cm} c}
\caption{presents factor relation to pull request decision.
First column lists factors in alphabet ascending order in each class, the rest the articles suggesting relations.\\
Horizontal Line in the middle of shape ($\ominus$) means the factor is removed when build models because of collinearity. \\
Filling: Filled ($\CIRCLE$) means \emph{significance is reported} and unfilled ($\Circle$) means \emph{significance is not reported because of not using statistical model or inconsistent conclusions}.\\
Size of filled shape: Big shape ($\CIRCLE$) shows \emph{statistically significant} relation and small shape ($\tiny \CIRCLE$) \emph{statistically insignificant} with 95\% confidence threshold.\\
Color: Blue {\color{blue} $\CIRCLE$} means a \emph{positive relation} (meaning increase in the chances of pull request acceptance or decreasing the latency of pull request), red {\color{red} $\CIRCLE$} means a \emph{negative relation}, gray {\color{gray} $\CIRCLE$} means \emph{uncertain relation} because of not using statistical model or no-linear conclusion.
\label{factor_table}}\\
\toprule%
 \centering%
 & {{\bfseries \cite{Gousios_exploratory}}}
 & {{\bfseries \cite{Tsay_influence}}}
 & {{\bfseries \cite{Yu_determinants}}}
 & {{\bfseries \cite{Baysal_investigating}}}
 & {{\bfseries \cite{Iyer_tse}}}
 & {{\bfseries \cite{Iyer_master}}}
 & {{\bfseries \cite{Rastogi_relationshipESEM}}}
 & {{\bfseries \cite{Khadke_predict}}}
 & {{\bfseries \cite{Weigerber_small}}}
 & {{\bfseries \cite{Kononenko_shopify}}}
 & {{\bfseries \cite{Zampetti_ci}}}
 & {{\bfseries \cite{Soares_acceptance}}}
 & {{\bfseries \cite{Soares_rejection}}}
 & {{\bfseries \cite{Golzadeh_discussion}}}
 & {{\bfseries \cite{Padhye_external}}}
 & {{\bfseries \cite{Legay_impact}}}
 & {{\bfseries \cite{Jiang_reopen}}}
 & {{\bfseries \cite{Rahman_insight}}}
 & {{\bfseries \cite{Baysal_firefox}}}
 & {{\bfseries \cite{Pinto_who}}}
 & {{\bfseries \cite{Bosu_reputation}}}
 & {{\bfseries \cite{Lee_onetime}}}
 & {{\bfseries \cite{Hechtl_coreness}}}
 & {{\bfseries \cite{Kovalenko_newcomer}}}
 & {{\bfseries \cite{Terrell_gender}}}
 & {{\bfseries \cite{Hilton_ci}}} \\
\endhead

\rowcolor[gray]{0.525}
\multicolumn{27}{c}{Developer Characteristics} \\

account\_creation\_days & & & & & & & \textcolor{red}{\normalsize $\CIRCLE$} & & & & & & & & & & & \textcolor{gray}{\normalsize $\Circle$} & & & & & & & & \\

\rowcolor[gray]{0.925}
agree\_diff & & & & & \textcolor{blue}{\normalsize $\CIRCLE$} & \textcolor{blue}{\normalsize $\CIRCLE$} & & & & & & & & & & & & & & & & & & & & \\

cons\_diff & & & & & \textcolor{blue}{\normalsize $\CIRCLE$} & \textcolor{blue}{\normalsize $\CIRCLE$} & & & & & & & & & & & & & & & & & & & & \\

\rowcolor[gray]{0.925}
contrib\_affiliation & & & & \textcolor{gray}{\normalsize $\Circle$} & & & & & & \textcolor{blue}{\normalsize $\CIRCLE$} & & & & & & & & & & & & & & & & \\

contrib\_agree & & & & & \textcolor{blue}{\tiny $\CIRCLE$} & \textcolor{blue}{\tiny $\CIRCLE$} & & & & & & & & & & & & & & & & & & & & \\

\rowcolor[gray]{0.925}
contrib\_cons & & & & & \textcolor{blue}{\normalsize $\CIRCLE$} & \textcolor{blue}{\normalsize $\CIRCLE$} & & & & & & & & & & & & & & & & & & & & \\

contrib\_country & & & & & & & \textcolor{gray}{\normalsize $\Circle$} & & & & & & & & & & & & & & & & & & & \\

\rowcolor[gray]{0.925}
contrib\_extra & & & & & \textcolor{red}{\normalsize $\CIRCLE$} & \textcolor{red}{\normalsize $\CIRCLE$} & & & & & & & & & & & & & & & & & & & & \\

contrib\_first\_emo & & & & & & \textcolor{blue}{\normalsize $\CIRCLE$} & & & & & & & & & & & & & & & & & & & & \\

\rowcolor[gray]{0.925}
contrib\_follow\_integrator & & \textcolor{blue}{\normalsize $\CIRCLE$} & & & \textcolor{blue}{\normalsize $\CIRCLE$} & \textcolor{blue}{\normalsize $\CIRCLE$} & \textcolor{blue}{\normalsize $\CIRCLE$} & & & & & & & & & & & & & & & & & & & \\

contrib\_gender & & & & & & & & & & & & & & & & & & & & & & & & & \textcolor{blue}{\normalsize $\CIRCLE$} & \\

\rowcolor[gray]{0.925}
contrib\_neur & & & & & \textcolor{red}{\tiny $\CIRCLE$} & \textcolor{red}{\tiny $\CIRCLE$} & & & & & & & & & & & & & & & & & & & & \\

contrib\_open & & & & & \textcolor{blue}{\normalsize $\CIRCLE$} & \textcolor{blue}{\normalsize $\CIRCLE$} & & & & & & & & & & & & & & & & & & & & \\

\rowcolor[gray]{0.925}
contrib\_rate\_author & & & & & & & & & & & \textcolor{gray}{\normalsize $\Circle$} & & & & & & & & & & & & & & & \\

core\_member & & \textcolor{blue}{\normalsize $\CIRCLE$} & \textcolor{blue}{\normalsize $\CIRCLE$} & & \textcolor{blue}{\normalsize $\CIRCLE$} & \textcolor{blue}{\normalsize $\CIRCLE$} & \textcolor{blue}{\tiny $\CIRCLE$} & & & & \textcolor{gray}{\normalsize $\Circle$} & \textcolor{blue}{\normalsize $\Circle$} & & & & & & & \textcolor{blue}{\normalsize $\Circle$} & \textcolor{blue}{\normalsize $\Circle$} & \textcolor{blue}{\normalsize $\CIRCLE$} & \textcolor{blue}{\normalsize $\CIRCLE$} & \textcolor{gray}{\normalsize $\Circle$} & & & \\

\rowcolor[gray]{0.925}
extra\_diff & & & & & \textcolor{blue}{\normalsize $\CIRCLE$} & \textcolor{blue}{\normalsize $\CIRCLE$} & & & & & & & & & & & & & & & & & & & & \\

first\_pr & & & & & & & & & & & & \textcolor{red}{\normalsize $\Circle$} & \textcolor{red}{\normalsize $\Circle$} & & & & & & & & & \textcolor{red}{\normalsize $\CIRCLE$} & & \textcolor{red}{\normalsize $\Circle$} & & \\

\rowcolor[gray]{0.925}
first\_response\_time & & & \textcolor{red}{\normalsize $\CIRCLE$} & & & & & & & & & & & & & & & & & & & & & & & \\

followers & & \textcolor{blue}{\normalsize $\CIRCLE$} & \textcolor{blue}{\normalsize $\CIRCLE$} & & \textcolor{gray}{\tiny $\CIRCLE$} & \textcolor{gray}{\tiny $\CIRCLE$} & \textcolor{blue}{\normalsize $\CIRCLE$} & & & & & & & & & & & & & & & & & & & \\

\rowcolor[gray]{0.925}
inte\_affiliation & & & & \textcolor{gray}{\normalsize $\Circle$} & & & & & & & & & & & & & & & & & & & & & & \\

inte\_agree & & & & & \textcolor{blue}{\tiny $\CIRCLE$} & \textcolor{blue}{\tiny $\CIRCLE$} & & & & & & & & & & & & & & & & & & & & \\

\rowcolor[gray]{0.925}
inte\_cons & & & & & \textcolor{blue}{\normalsize $\CIRCLE$} & \textcolor{blue}{\normalsize $\CIRCLE$} & & & & & & & & & & & & & & & & & & & & \\

inte\_extra & & & & & \textcolor{blue}{\normalsize $\CIRCLE$} & \textcolor{blue}{\normalsize $\CIRCLE$} & & & & & & & & & & & & & & & & & & & & \\

\rowcolor[gray]{0.925}
inte\_first\_emo & & & & & & \textcolor{blue}{\normalsize $\CIRCLE$} & & & & & & & & & & & & & & & & & & & & \\

inte\_neur & & & & & \textcolor{blue}{\normalsize $\CIRCLE$} & \textcolor{blue}{\normalsize $\CIRCLE$} & & & & & & & & & & & & & & & & & & & & \\

\rowcolor[gray]{0.925}
inte\_open & & & & & \textcolor{blue}{\normalsize $\CIRCLE$} & \textcolor{blue}{\normalsize $\CIRCLE$} & & & & & & & & & & & & & & & & & & & & \\

neur\_diff & & & & & \textcolor{blue}{\normalsize $\CIRCLE$} & \textcolor{blue}{\normalsize $\CIRCLE$} & & & & & & & & & & & & & & & & & & & & \\

\rowcolor[gray]{0.925}
open\_diff & & & & & \textcolor{blue}{\tiny $\CIRCLE$} & \textcolor{blue}{\tiny $\CIRCLE$} & & & & & & & & & & & & & & & & & & & & \\

perc\_contrib\_neg\_emo & & & & & & \textcolor{red}{\normalsize $\CIRCLE$} & & & & & & & & & & & & & & & & & & & & \\

\rowcolor[gray]{0.925}
perc\_contrib\_pos\_emo & & & & & & \textcolor{blue}{\normalsize $\CIRCLE$} & & & & & & & & & & & & & & & & & & & & \\

perc\_inte\_neg\_emo & & & & & & \textcolor{red}{\normalsize $\CIRCLE$} & & & & & & & & & & & & & & & & & & & & \\

\rowcolor[gray]{0.925}
perc\_inte\_pos\_emo & & & & & & \textcolor{blue}{\normalsize $\CIRCLE$} & & & & & & & & & & & & & & & & & & & & \\

prev\_pullreqs & \textcolor{gray}{\normalsize $\Circle$} & & & \textcolor{gray}{\normalsize $\CIRCLE$} & & & \textcolor{blue}{\normalsize $\CIRCLE$} & & & \textcolor{blue}{\normalsize $\CIRCLE$} & & & & & & \textcolor{blue}{\normalsize $\Circle$} & & & & & & & & & & \\

\rowcolor[gray]{0.925}
prior\_interaction & & \textcolor{blue}{\normalsize $\CIRCLE$} & & & \textcolor{blue}{\normalsize $\CIRCLE$} & \textcolor{blue}{\normalsize $\CIRCLE$} & & & & & & & & & & & & & & & & & & & & \\

prior\_review\_num & & & & \textcolor{gray}{\tiny $\CIRCLE$} & & & & & & & & & & & & & & & & & & & & & & \\

\rowcolor[gray]{0.925}
same\_affiliation & & & & \textcolor{gray}{\normalsize $\Circle$} & & & & & & & & & & & & & & & & & & & & & & \\

same\_country & & & & & & & \textcolor{blue}{\normalsize $\CIRCLE$} & & & & & & & & & & & & & & & & & & & \\

\rowcolor[gray]{0.925}
social\_strength & & & \textcolor{blue}{\normalsize $\CIRCLE$} & & & & & & & & & & & & & & & & & & & & & & & \\

\rowcolor[gray]{0.525}
\multicolumn{27}{c}{Project Characteristics} \\

asserts\_per\_kloc & \textcolor{gray}{\normalsize $\Circle$} & & & & & & \textcolor{gray}{\normalsize $\ominus$} & & & & & & & & & & & & & & & & & & & \\

\rowcolor[gray]{0.925}
fork\_num & & \textcolor{gray}{\normalsize $\ominus$} & & & & & & \textcolor{red}{\normalsize $\Circle$} & & & & & & & & & & \textcolor{gray}{\normalsize $\Circle$} & & & & & & & & \\

integrator\_availability & & & \textcolor{red}{\normalsize $\CIRCLE$} & & & & & & & & & & & & & & & & & & & & & & & \\

\rowcolor[gray]{0.925}
language & & & & & & & & & & & & \textcolor{gray}{\normalsize $\Circle$} & & & & & & \textcolor{gray}{\normalsize $\Circle$} & & & & & & & & \\

open\_issue\_num & & & & & & & & \textcolor{blue}{\normalsize $\Circle$} & & & & & & & & & & & & & & & & & & \\

\rowcolor[gray]{0.925}
open\_pr\_num & & & \textcolor{red}{\normalsize $\CIRCLE$} & \textcolor{gray}{\normalsize $\Circle$} & & & & & & & & & & & & & & & & & & & & & & \\

perc\_external\_contribs & \textcolor{gray}{\normalsize $\Circle$} & & & & & & \textcolor{red}{\normalsize $\CIRCLE$} & & & & & & & & & & & & & & & & & & & \\

\rowcolor[gray]{0.925}
project\_age & & \textcolor{red}{\normalsize $\CIRCLE$} & \textcolor{blue}{\normalsize $\CIRCLE$} & & \textcolor{red}{\normalsize $\CIRCLE$} & \textcolor{red}{\normalsize $\CIRCLE$} & \textcolor{red}{\normalsize $\CIRCLE$} & & & & & & & & & & & \textcolor{gray}{\normalsize $\Circle$} & & & & & & & & \\

pr\_succ\_rate & & & & & & & & \textcolor{blue}{\normalsize $\Circle$} & & & & & & & & & & & & & & & & & & \\

\rowcolor[gray]{0.925}
pushed\_delta & & & & & & & & \textcolor{red}{\normalsize $\Circle$} & & & & & & & & & & & & & & & & & & \\

requester\_succ\_rate & \textcolor{gray}{\normalsize $\Circle$} & & & & & & \textcolor{blue}{\normalsize $\CIRCLE$} & \textcolor{blue}{\normalsize $\Circle$} & & & \textcolor{gray}{\normalsize $\Circle$} & & & & & & & & & & & & & & & \\

\rowcolor[gray]{0.925}
sloc & \textcolor{gray}{\normalsize $\Circle$} & & & & & & \textcolor{red}{\normalsize $\CIRCLE$} & & & & & & & & & & & & & & & & & & & \\

stars & & \textcolor{red}{\normalsize $\CIRCLE$} & & & \textcolor{red}{\normalsize $\CIRCLE$} & \textcolor{red}{\normalsize $\CIRCLE$} & \textcolor{red}{\normalsize $\CIRCLE$} & \textcolor{red}{\normalsize $\Circle$} & & & & & & & & & & & & & & & & & & \\

\rowcolor[gray]{0.925}
team\_size & \textcolor{gray}{\normalsize $\Circle$} & \textcolor{red}{\normalsize $\CIRCLE$} & \textcolor{blue}{\tiny $\CIRCLE$} & & \textcolor{red}{\tiny $\CIRCLE$} & \textcolor{red}{\tiny $\CIRCLE$} & \textcolor{gray}{\normalsize $\ominus$} & \textcolor{blue}{\normalsize $\Circle$} & & & & & & & & & & \textcolor{gray}{\normalsize $\Circle$} & & & & & & & & \\

test\_cases\_per\_kloc & \textcolor{gray}{\normalsize $\ominus$} & & & & & & \textcolor{gray}{\normalsize $\ominus$} & & & & & & & & & & & & & & & & & & & \\

\rowcolor[gray]{0.925}
test\_lines\_per\_kloc & \textcolor{gray}{\normalsize $\Circle$} & & & & & & \textcolor{blue}{\normalsize $\CIRCLE$} & & & & & & & & & & & & & & & & & & & \\

\rowcolor[gray]{0.525}
\multicolumn{27}{c}{Pull Request Characteristics} \\

at\_tag & & & \textcolor{blue}{\tiny $\CIRCLE$} & & & & & & & & & & & & & & & & & & & & & & & \\

\rowcolor[gray]{0.925}
bug\_fix & & & & & & & & & & & \textcolor{gray}{\normalsize $\Circle$} & & & & \textcolor{blue}{\normalsize $\Circle$} & & & & & & & & & & & \\

churn\_addition & & & \textcolor{red}{\normalsize $\CIRCLE$} & & & & & & & & & & & & & & & & & & & & & & & \\

\rowcolor[gray]{0.925}
churn\_deletion & & & \textcolor{blue}{\normalsize $\CIRCLE$} & & & & & & & & \textcolor{gray}{\normalsize $\Circle$} & & & & & & & & & & & & & & & \\

ci\_build\_num & & & & & & & & & & & \textcolor{gray}{\normalsize $\Circle$} & & & & & & & & & & & & & & & \\

\rowcolor[gray]{0.925}
ci\_exists & & & & & & & & & & & & & & & & & & & & & & & & & & \textcolor{red}{\normalsize $\CIRCLE$} \\

ci\_failed\_perc & & & & & & & & & & & \textcolor{gray}{\normalsize $\Circle$} & & & & & & & & & & & & & & & \\

\rowcolor[gray]{0.925}
ci\_first\_build\_status & & & & & & & & & & & \textcolor{blue}{\normalsize $\CIRCLE$} & & & & & & & & & & & & & & & \\

ci\_last\_build\_status & & & & & & & & & & & \textcolor{blue}{\normalsize $\CIRCLE$} & & & & & & & & & & & & & & & \\

\rowcolor[gray]{0.925}
ci\_latency & & & \textcolor{blue}{\normalsize $\CIRCLE$} & & & & & & & & & & & & & & & & & & & & & & & \\

ci\_test\_passed & & & \textcolor{blue}{\normalsize $\CIRCLE$} & & & & & & & & & & & & & & & & & & & & & & & \\

\rowcolor[gray]{0.925}
comment\_conflict & \textcolor{gray}{\normalsize $\Circle$} & & & & & & & & & & & & & & & & & & & & & & & & & \\

commits\_on\_files\_touched & \textcolor{gray}{\normalsize $\Circle$} & & \textcolor{blue}{\normalsize $\CIRCLE$} & & & & & & & & & & & & & & & & & & & & & & & \\

\rowcolor[gray]{0.925}
contrib\_comment & & & & & & & & & & & & & & \textcolor{red}{\normalsize $\Circle$} & & & & & & & & & & & & \\

description\_length & & & \textcolor{red}{\normalsize $\CIRCLE$} & & & & & & & & & & & & & & & & & & & & & & & \\

\rowcolor[gray]{0.925}
files\_added & & & & & & & & & & & & \textcolor{red}{\normalsize $\Circle$} & & & & & & & & & & & & & & \\

files\_changed & \textcolor{gray}{\normalsize $\Circle$} & \textcolor{red}{\normalsize $\CIRCLE$} & & & \textcolor{gray}{\normalsize $\ominus$} & \textcolor{gray}{\normalsize $\ominus$} & \textcolor{blue}{\tiny $\CIRCLE$} & \textcolor{blue}{\normalsize $\Circle$} & & \textcolor{gray}{\normalsize $\ominus$} & \textcolor{gray}{\normalsize $\Circle$} & \textcolor{gray}{\normalsize $\Circle$} & \textcolor{red}{\normalsize $\Circle$} & & & & & & & & & & & & & \\

\rowcolor[gray]{0.925}
files\_deleted & & & & & & & & & & & & \textcolor{blue}{\normalsize $\Circle$} & & & & & & & & & & & & & & \\

friday\_effect & & & \textcolor{blue}{\tiny $\CIRCLE$} & & & & & & & & & & & & & & & & & & & & & & & \\

\rowcolor[gray]{0.925}
has\_comments & & & & & & & & & & & & & \textcolor{red}{\normalsize $\Circle$} & \textcolor{red}{\normalsize $\Circle$} & & & & & & & & & & & & \\

has\_exchange & & & & & & & & & & & & & & \textcolor{red}{\normalsize $\Circle$} & & & & & & & & & & & & \\

\rowcolor[gray]{0.925}
hash\_tag & \textcolor{gray}{\normalsize $\Circle$} & & \textcolor{blue}{\tiny $\CIRCLE$} & & & & & & & & & & & & & & & & & & & & & & & \\

has\_participants & & & & & & & & & & & & & & \textcolor{red}{\normalsize $\Circle$} & & & & & & & & & & & & \\

\rowcolor[gray]{0.925}
inte\_comment & & & & & & & & & & & & & & \textcolor{blue}{\normalsize $\Circle$} & & & & & & & & & & & & \\

lifetime\_minutes & & & & & & & & & & & \textcolor{gray}{\normalsize $\Circle$} & \textcolor{red}{\normalsize $\Circle$} & & & & \textcolor{red}{\normalsize $\Circle$} & & & & & & & & & & \\

\rowcolor[gray]{0.925}
num\_code\_comments & & & & & & & & & & \textcolor{gray}{\normalsize $\ominus$} & \textcolor{gray}{\normalsize $\Circle$} & & & & & & & & & & & & & & & \\

num\_code\_comments\_con & & & & & & & & & & \textcolor{gray}{\normalsize $\ominus$} & & & & & & & & & & & & & & & & \\

\rowcolor[gray]{0.925}
num\_comments & \textcolor{gray}{\normalsize $\Circle$} & \textcolor{red}{\normalsize $\CIRCLE$} & \textcolor{red}{\normalsize $\CIRCLE$} & & \textcolor{red}{\normalsize $\CIRCLE$} & \textcolor{red}{\normalsize $\CIRCLE$} & \textcolor{red}{\normalsize $\CIRCLE$} & & & \textcolor{gray}{\normalsize $\ominus$} & \textcolor{gray}{\normalsize $\Circle$} & & & & & & & & & & & & & & & \\

num\_comments\_con & & & & & & & & & & \textcolor{gray}{\tiny $\CIRCLE$} & & & & & & & & & & & & & & & & \\

\rowcolor[gray]{0.925}
num\_commits & \textcolor{gray}{\normalsize $\Circle$} & & \textcolor{blue}{\normalsize $\CIRCLE$} & & & & & \textcolor{red}{\normalsize $\Circle$} & & \textcolor{gray}{\normalsize $\ominus$} & \textcolor{gray}{\normalsize $\Circle$} & \textcolor{red}{\normalsize $\Circle$} & \textcolor{red}{\normalsize $\Circle$} & & & & & & & & & & & & & \\

num\_participants & \textcolor{gray}{\normalsize $\Circle$} & & & & & & & & & \textcolor{red}{\normalsize $\CIRCLE$} & & & & & & & & & & & & & & & & \\

\rowcolor[gray]{0.925}
other\_comment & & & & & & & & & & & & & & \textcolor{red}{\normalsize $\Circle$} & & & & & & & & & & & & \\

part\_num\_code & & & & & & & & & & \textcolor{blue}{\normalsize $\CIRCLE$} & & & & & & & & & & & & & & & & \\

\rowcolor[gray]{0.925}
perc\_neg\_emotion & & & & & & \textcolor{red}{\normalsize $\CIRCLE$} & & & & & & & & & & & & & & & & & & & & \\

perc\_pos\_emotion & & & & & & \textcolor{blue}{\normalsize $\CIRCLE$} & & & & & & & & & & & & & & & & & & & & \\

\rowcolor[gray]{0.925}
reopen\_or\_not & & & & & & & & & & & & & & & & & \textcolor{red}{\normalsize $\Circle$} & & & & & & & & & \\

core\_comment & & & & & & & & & & & & & & \textcolor{red}{\normalsize $\Circle$} & & & & & & & & & & & & \\

\rowcolor[gray]{0.925}
src\_churn & \textcolor{gray}{\normalsize $\Circle$} & \textcolor{red}{\normalsize $\CIRCLE$} & & \textcolor{gray}{\tiny $\CIRCLE$} & \textcolor{red}{\normalsize $\CIRCLE$} & \textcolor{red}{\normalsize $\CIRCLE$} & \textcolor{red}{\normalsize $\CIRCLE$} & \textcolor{red}{\normalsize $\Circle$} & \textcolor{red}{\normalsize $\Circle$} & \textcolor{red}{\normalsize $\CIRCLE$} & & & & & & & & & & & & & & & & \\

test\_churn & \textcolor{gray}{\normalsize $\Circle$} & & & & & & & & & & & & & & & & & & & & & & & & & \\

\rowcolor[gray]{0.925}
test\_inclusion & & \textcolor{blue}{\normalsize $\CIRCLE$} & \textcolor{blue}{\normalsize $\CIRCLE$} & & \textcolor{blue}{\normalsize $\CIRCLE$} & \textcolor{blue}{\normalsize $\CIRCLE$} & \textcolor{blue}{\normalsize $\CIRCLE$} & & & & \textcolor{gray}{\normalsize $\Circle$} & & & & & & & & & & & & & & & \\

\bottomrule

\end{longtable}
\end{landscape}
\normalsize
\twocolumn
\clearpage

\appendices
\onecolumn
\begin{landscape}
\section{Results of different contexts}
\label{appendix-result-different-contexts}
\begin{table*}[!h] \centering
  \setlength\tabcolsep{4pt}
  \tiny
  \caption{presents the full results in different contexts} 
  \label{result-different-contexts} 
\begin{tabular}{ lcccccccccccc} 
\\[-4.9ex]
\toprule
\\[-2.8ex] 
 & \multicolumn{12}{c}{\textit{Dependent variable: merged\_or\_not=1}} \\ 
\cline{2-13} 
\\[-1.8ex] 
& \multicolumn{2}{c}{same user or not} & \multicolumn{2}{c}{has comments or not} & \multicolumn{2}{c}{ci exists or not} & \multicolumn{3}{c}{different team sizes} & \multicolumn{3}{c}{different periods}\\
& yes & no & yes & no & yes & no & small & mid & large & before 2016.6 & 2016.6-2018.6 & after 2018.6 \\[-0.5ex] 
\cmidrule(lr){2-3} \cmidrule(lr){4-5} \cmidrule(lr){6-7} \cmidrule(lr){8-10} \cmidrule(lr){11-13}\\[-3.2ex] 
(Intercept) & $10.4^{***}$ & $34.4^{***}$ & $13.1^{***}$ & $42.4^{***}$ & $20.4^{***}$ & $13.3^{***}$ & $24.9^{***}$ & $20.7^{***}$ & $15.9^{***}$ & $6.9^{***}$ & $16.6^{***}$ & $7.1^{***}$ \\
 
 prior\_review\_num & \ccell{cellgray}{$2.86^{***}[31.17]$} & \ccell{cellgray}{$0.98^{***}[\,\,\,0.04]$} & \ccell{cellgray}{$1.51^{***}[14.40]$} & \ccell{cellgray}{$1.91^{***}[22.12]$} & $1.53^{***}[13.76]$ & $1.53^{***}[11.95]$ & \ccell{cellgray}{$1.59^{***}[11.27]$} & \ccell{cellgray}{$1.41^{***}[\,\,\,8.80]$} & \ccell{cellgray}{$1.57^{***}[18.89]$} & \ccell{cellgray}{$1.30^{***}[\,\,\,6.25]$} & \ccell{cellgray}{$1.63^{***}[14.12]$} & \ccell{cellgray}{$1.72^{***}[17.00]$} \\

 lifetime\_minutes & \ccell{cellgray}{$0.66^{***}[19.09]$} & \ccell{cellgray}{$0.52^{***}[43.67]$} & \ccell{cellgray}{$0.61^{***}[29.79]$} & \ccell{cellgray}{$0.70^{***}[12.97]$} & $0.60^{***}[21.52]$ & $0.61^{***}[20.78]$ & $0.54^{***}[24.47]$ & $0.61^{***}[20.16]$ & $0.67^{***}[16.55]$ & \ccell{cellgray}{$0.65^{***}[20.26]$} & \ccell{cellgray}{$0.57^{***}[20.83]$} & \ccell{cellgray}{$0.62^{***}[12.69]$} \\

 core\_member & \ccell{cellgray}{$1.26^{***}[\,\,\,9.36]$} & \ccell{cellgray}{$1.13^{***}[\,\,\,1.31]$} & $1.29^{***}[\,\,\,5.55]$ & $1.33^{***}[\,\,\,5.85]$ & $1.30^{***}[\,\,\,5.28]$ & $1.26^{***}[\,\,\,4.66]$ & $1.42^{***}[\,\,\,5.90]$ & $1.28^{***}[\,\,\,5.24]$ & $1.19^{***}[\,\,\,3.24]$ & $1.27^{***}[\,\,\,5.99]$ & $1.34^{***}[\,\,\,4.94]$ & $1.29^{***}[\,\,\,3.14]$ \\

 prev\_pullreqs & $0.61^{***}[\,\,\,5.85]$ & $1.17^{***}[\,\,\,1.24]$ & $1.13^{***}[\,\,\,0.69]$ & $0.95^{***}[\,\,\,0.09]$ & $1.14^{***}[\,\,\,0.64]$ & $1.12^{***}[\,\,\,0.56]$ & $1.21^{***}[\,\,\,0.79]$ & $1.21^{***}[\,\,\,1.37]$ & $1.13^{***}[\,\,\,0.84]$ & $1.08^{***}[\,\,\,0.29]$ & $1.26^{***}[\,\,\,1.59]$ & $1.24^{***}[\,\,\,1.32]$ \\
 
 num\_commits & \ccell{cellgray}{$1.23^{***}[\,\,\,3.78]$} & \ccell{cellgray}{$1.46^{***}[10.43]$} & \ccell{cellgray}{$1.49^{***}[11.33]$} & \ccell{cellgray}{$0.98^{**\,\,\,}[\,\,\,0.03]$} & $1.32^{***}[\,\,\,4.85]$ & $1.25^{***}[\,\,\,3.57]$ & $1.36^{***}[\,\,\,4.64]$ & $1.31^{***}[\,\,\,4.53]$ & $1.26^{***}[\,\,\,4.36]$ & $1.18^{***}[\,\,\,2.31]$ & $1.32^{***}[\,\,\,4.13]$ & $1.36^{***}[\,\,\,4.84]$ \\

 hash\_tag & $1.14^{***}[\,\,\,2.52]$ & $1.10^{***}[\,\,\,1.34]$ & $1.14^{***}[\,\,\,2.27]$ & $1.09^{***}[\,\,\,0.65]$ & $1.12^{***}[\,\,\,1.46]$ & $1.10^{***}[\,\,\,1.14]$ & $1.13^{***}[\,\,\,1.37]$ & $1.11^{***}[\,\,\,1.16]$ & $1.11^{***}[\,\,\,1.38]$ & $1.04^{***}[\,\,\,0.22]$ & $1.13^{***}[\,\,\,1.44]$ & $1.13^{***}[\,\,\,1.30]$ \\

 first\_pr & $0.91^{***}[\,\,\,1.82]$ & $0.96^{***}[\,\,\,0.30]$ & $0.95^{***}[\,\,\,0.32]$ & $0.95^{***}[\,\,\,0.31]$ & $0.94^{***}[\,\,\,0.45]$ & $0.96^{***}[\,\,\,0.27]$ & $0.95^{***}[\,\,\,0.23]$ & $0.97^{***}[\,\,\,0.15]$ & $0.96^{***}[\,\,\,0.31]$ & $0.95^{***}[\,\,\,0.50]$ & $0.96^{***}[\,\,\,0.20]$ & $0.94^{***}[\,\,\,0.32]$ \\

 files\_added & $0.88^{***}[\,\,\,1.57]$ & $0.95^{***}[\,\,\,0.30]$ & $0.90^{***}[\,\,\,0.83]$ & $0.92^{***}[\,\,\,0.57]$ & $0.90^{***}[\,\,\,0.90]$ & $0.92^{***}[\,\,\,0.53]$ & $0.90^{***}[\,\,\,0.52]$ & $0.89^{***}[\,\,\,1.04]$ & $0.90^{***}[\,\,\,0.94]$ & $0.97^{***}[\,\,\,0.11]$ & $0.88^{***}[\,\,\,0.95]$ & $0.86^{***}[\,\,\,1.34]$ \\
 
 reopen\_or\_not & $0.93^{***}[\,\,\,1.52]$ & $0.99^{.\,\,\,\,\,\,\,}[\,\,\,0.01]$ & $0.97^{***}[\,\,\,0.17]$ & $0.92^{***}[\,\,\,2.39]$ & $0.96^{***}[\,\,\,0.40]$ & $0.97^{***}[\,\,\,0.24]$ & $0.97^{***}[\,\,\,0.23]$ & $0.96^{***}[\,\,\,0.48]$ & $0.96^{***}[\,\,\,0.43]$ & $0.99^{***}[\,\,\,0.03]$ & $0.97^{***}[\,\,\,0.18]$ & $0.93^{***}[\,\,\,1.06]$ \\

 open\_pr\_num & $0.73^{***}[\,\,\,1.37]$ & $1.06^{***}[\,\,\,0.05]$ & $0.92^{***}[\,\,\,0.09]$ & $0.60^{***}[\,\,\,3.07]$ & $0.83^{***}[\,\,\,0.33]$ & $0.76^{***}[\,\,\,0.93]$ & $0.81^{***}[\,\,\,1.26]$ & $0.83^{***}[\,\,\,0.81]$ & $0.98^{\,\,\,\,\,\,\,\,\,}[\,\,\,0.01]$ & $0.77^{***}[\,\,\,1.34]$ & $0.72^{***}[\,\,\,0.53]$ & $0.75^{***}[\,\,\,0.28]$ \\

 contrib\_open & $1.13^{***}[\,\,\,1.28]$ & $1.06^{***}[\,\,\,0.38]$ & $1.05^{***}[\,\,\,0.25]$ & $1.09^{***}[\,\,\,0.24]$ & $1.05^{***}[\,\,\,0.15]$ & $1.07^{***}[\,\,\,0.34]$ & $1.05^{***}[\,\,\,0.12]$ & $1.12^{***}[\,\,\,0.70]$ & $1.05^{***}[\,\,\,0.25]$ & $1.07^{***}[\,\,\,0.46]$ & $1.07^{***}[\,\,\,0.27]$ & $1.02^{**\,\,\,}[\,\,\,0.02]$ \\

 description\_length & $1.06^{***}[\,\,\,0.45]$ & $1.02^{***}[\,\,\,0.05]$ & $1.01^{***}[\,\,\,0.03]$ & $1.12^{***}[\,\,\,1.11]$ & $1.04^{***}[\,\,\,0.17]$ & $1.04^{***}[\,\,\,0.13]$ & $1.03^{***}[\,\,\,0.08]$ & $1.03^{***}[\,\,\,0.07]$ & $1.05^{***}[\,\,\,0.29]$ & $0.99^{*\,\,\,\,\,\,}[\,\,\,0.01]$ & $1.06^{***}[\,\,\,0.26]$ & $1.07^{***}[\,\,\,0.36]$ \\

 commits\_on\_files\_touched & $1.06^{***}[\,\,\,0.41]$ & $1.11^{***}[\,\,\,1.16]$ & $1.05^{***}[\,\,\,0.23]$ & $1.18^{***}[\,\,\,1.86]$ & $1.06^{***}[\,\,\,0.24]$ & $1.13^{***}[\,\,\,1.39]$ & $1.10^{***}[\,\,\,0.61]$ & $1.12^{***}[\,\,\,0.90]$ & $1.05^{***}[\,\,\,0.20]$ & \ccell{cellgray}{$1.30^{***}[\,\,\,7.25]$} & \ccell{cellgray}{$0.99^{\,\,\,\,\,\,\,\,\,}[\,\,\,0.00]$} & \ccell{cellgray}{$0.99^{\,\,\,\,\,\,\,\,\,}[\,\,\,0.00]$} \\
stars & $0.83^{***}[\,\,\,0.40]$ & $0.94^{***}[\,\,\,0.05]$ & $0.83^{***}[\,\,\,0.39]$ & $0.83^{***}[\,\,\,0.37]$ & $0.85^{***}[\,\,\,0.24]$ & $0.75^{***}[\,\,\,0.81]$ & $0.81^{***}[\,\,\,0.69]$ & $0.89^{***}[\,\,\,0.17]$ & $1.05^{**\,\,\,}[\,\,\,0.01]$ & $0.83^{***}[\,\,\,0.45]$ & $0.72^{***}[\,\,\,0.44]$ & $0.71^{***}[\,\,\,0.50]$ \\
project\_age & $1.08^{***}[\,\,\,0.19]$ & $1.24^{***}[\,\,\,1.37]$ & $1.08^{***}[\,\,\,0.16]$ & $1.16^{***}[\,\,\,0.53]$ & $1.10^{***}[\,\,\,0.20]$ & $1.15^{***}[\,\,\,0.50]$ & $1.04^{***}[\,\,\,0.05]$ & $1.08^{***}[\,\,\,0.15]$ & $0.98^{*\,\,\,\,\,\,}[\,\,\,0.01]$ & $0.89^{***}[\,\,\,0.40]$ & $1.69^{***}[\,\,\,2.27]$ & $3.80^{***}[\,\,\,4.81]$ \\
files\_changed & $0.94^{***}[\,\,\,0.18]$ & $0.90^{***}[\,\,\,0.54]$ & $0.95^{***}[\,\,\,0.11]$ & $0.90^{***}[\,\,\,0.43]$ & $0.94^{***}[\,\,\,0.15]$ & $0.90^{***}[\,\,\,0.47]$ & $0.93^{***}[\,\,\,0.16]$ & $0.91^{***}[\,\,\,0.33]$ & $0.93^{***}[\,\,\,0.21]$ & $0.86^{***}[\,\,\,1.13]$ & $0.95^{***}[\,\,\,0.09]$ & $0.94^{***}[\,\,\,0.13]$ \\
test\_churn & $1.05^{***}[\,\,\,0.15]$ & $1.09^{***}[\,\,\,0.52]$ & $1.08^{***}[\,\,\,0.39]$ & $0.99^{\,\,\,\,\,\,\,\,\,}[\,\,\,0.01]$ & $1.07^{***}[\,\,\,0.27]$ & $1.05^{***}[\,\,\,0.15]$ & $1.11^{***}[\,\,\,0.45]$ & $1.07^{***}[\,\,\,0.25]$ & $1.04^{***}[\,\,\,0.11]$ & $1.07^{***}[\,\,\,0.34]$ & $1.10^{***}[\,\,\,0.40]$ & $1.08^{***}[\,\,\,0.23]$ \\
account\_creation\_days & $1.03^{***}[\,\,\,0.13]$ & $1.11^{***}[\,\,\,1.70]$ & $1.05^{***}[\,\,\,0.28]$ & $1.17^{***}[\,\,\,2.26]$ & $1.08^{***}[\,\,\,0.58]$ & $1.04^{***}[\,\,\,0.22]$ & $1.06^{***}[\,\,\,0.33]$ & $1.11^{***}[\,\,\,1.04]$ & $1.02^{***}[\,\,\,0.06]$ & $0.99^{*\,\,\,\,\,\,}[\,\,\,0.01]$ & $1.02^{***}[\,\,\,0.02]$ & $1.03^{***}[\,\,\,0.06]$ \\
team\_size & $0.94^{***}[\,\,\,0.08]$ & $1.09^{***}[\,\,\,0.19]$ & $1.06^{***}[\,\,\,0.07]$ & $0.92^{***}[\,\,\,0.14]$ & $1.02^{*\,\,\,\,\,\,}[\,\,\,0.00]$ & $0.96^{*\,\,\,\,\,\,}[\,\,\,0.02]$ & $1.06^{***}[\,\,\,0.30]$ & $1.00^{\,\,\,\,\,\,\,\,\,}[\,\,\,0.00]$ & $0.96^{***}[\,\,\,0.04]$ & $0.88^{***}[\,\,\,0.37]$ & $1.09^{***}[\,\,\,0.07]$ & $0.85^{***}[\,\,\,0.16]$ \\
pushed\_delta & $1.02^{***}[\,\,\,0.07]$ & $1.06^{***}[\,\,\,0.46]$ & $1.03^{***}[\,\,\,0.15]$ & $1.06^{***}[\,\,\,0.38]$ & $1.04^{***}[\,\,\,0.17]$ & $1.04^{***}[\,\,\,0.18]$ & $1.06^{***}[\,\,\,0.31]$ & $1.05^{***}[\,\,\,0.25]$ & $1.02^{***}[\,\,\,0.08]$ & $1.04^{***}[\,\,\,0.26]$ & $1.03^{***}[\,\,\,0.05]$ & $1.04^{***}[\,\,\,0.12]$ \\
integrator\_availability & $0.98^{***}[\,\,\,0.07]$ & $1.03^{***}[\,\,\,0.15]$ & $1.00^{\,\,\,\,\,\,\,\,\,}[\,\,\,0.00]$ & $0.99^{\,\,\,\,\,\,\,\,\,}[\,\,\,0.01]$ & $0.99^{***}[\,\,\,0.03]$ & $1.01^{*\,\,\,\,\,\,}[\,\,\,0.03]$ & $1.03^{***}[\,\,\,0.07]$ & $1.01^{\,\,\,\,\,\,\,\,\,}[\,\,\,0.00]$ & $0.97^{***}[\,\,\,0.14]$ & $1.00^{\,\,\,\,\,\,\,\,\,}[\,\,\,0.00]$ & $0.97^{***}[\,\,\,0.06]$ & $0.98^{**\,\,\,}[\,\,\,0.03]$ \\
test\_inclusion & $1.03^{***}[\,\,\,0.06]$ & $1.00^{\,\,\,\,\,\,\,\,\,}[\,\,\,0.00]$ & $1.02^{***}[\,\,\,0.02]$ & $1.02^{*\,\,\,\,\,\,}[\,\,\,0.02]$ & $1.03^{***}[\,\,\,0.05]$ & $0.97^{***}[\,\,\,0.07]$ & $1.00^{\,\,\,\,\,\,\,\,\,}[\,\,\,0.00]$ & $1.03^{***}[\,\,\,0.05]$ & $1.01^{**\,\,\,}[\,\,\,0.02]$ & $1.00^{\,\,\,\,\,\,\,\,\,}[\,\,\,0.00]$ & $1.01^{*\,\,\,\,\,\,}[\,\,\,0.01]$ & $1.01^{\,\,\,\,\,\,\,\,\,}[\,\,\,0.00]$ \\
contrib\_neur & $1.02^{***}[\,\,\,0.04]$ & $1.05^{***}[\,\,\,0.27]$ & $1.01^{*\,\,\,\,\,\,}[\,\,\,0.01]$ & $1.04^{***}[\,\,\,0.05]$ & $1.01^{.\,\,\,\,\,\,\,}[\,\,\,0.00]$ & $1.03^{**\,\,\,}[\,\,\,0.05]$ & $1.00^{\,\,\,\,\,\,\,\,\,}[\,\,\,0.00]$ & $1.07^{***}[\,\,\,0.25]$ & $0.99^{**\,\,\,}[\,\,\,0.02]$ & $1.02^{***}[\,\,\,0.03]$ & $1.02^{***}[\,\,\,0.03]$ & $0.99^{\,\,\,\,\,\,\,\,\,}[\,\,\,0.00]$ \\
contrib\_cons & $1.02^{***}[\,\,\,0.04]$ & $1.04^{***}[\,\,\,0.17]$ & $1.05^{***}[\,\,\,0.20]$ & $0.94^{***}[\,\,\,0.12]$ & $1.03^{***}[\,\,\,0.05]$ & $1.02^{**\,\,\,}[\,\,\,0.04]$ & $1.05^{***}[\,\,\,0.12]$ & $1.03^{***}[\,\,\,0.04]$ & $1.03^{***}[\,\,\,0.07]$ & $1.01^{**\,\,\,}[\,\,\,0.02]$ & $1.03^{***}[\,\,\,0.05]$ & $1.12^{***}[\,\,\,0.54]$ \\
contrib\_gender & $0.98^{***}[\,\,\,0.04]$ & $0.97^{***}[\,\,\,0.13]$ & $0.97^{***}[\,\,\,0.10]$ & $0.99^{\,\,\,\,\,\,\,\,\,}[\,\,\,0.00]$ & $0.98^{***}[\,\,\,0.06]$ & $0.98^{**\,\,\,}[\,\,\,0.04]$ & $0.97^{***}[\,\,\,0.08]$ & $0.97^{***}[\,\,\,0.09]$ & $0.99^{**\,\,\,}[\,\,\,0.02]$ & $0.99^{\,\,\,\,\,\,\,\,\,}[\,\,\,0.01]$ & $0.97^{***}[\,\,\,0.10]$ & $1.01^{\,\,\,\,\,\,\,\,\,}[\,\,\,0.00]$ \\
pr\_succ\_rate & $0.98^{***}[\,\,\,0.04]$ & $0.99^{*\,\,\,\,\,\,}[\,\,\,0.01]$ & $0.98^{***}[\,\,\,0.03]$ & $0.97^{***}[\,\,\,0.05]$ & $0.97^{***}[\,\,\,0.06]$ & $1.02^{**\,\,\,}[\,\,\,0.04]$ & $0.94^{***}[\,\,\,0.25]$ & $0.99^{\,\,\,\,\,\,\,\,\,}[\,\,\,0.01]$ & $0.96^{***}[\,\,\,0.10]$ & $0.97^{***}[\,\,\,0.14]$ & $1.00^{\,\,\,\,\,\,\,\,\,}[\,\,\,0.00]$ & $1.11^{***}[\,\,\,0.13]$ \\
contrib\_agree & $0.98^{***}[\,\,\,0.04]$ & $0.99^{*\,\,\,\,\,\,}[\,\,\,0.01]$ & $0.98^{***}[\,\,\,0.03]$ & $0.96^{**\,\,\,}[\,\,\,0.04]$ & $0.99^{**\,\,\,}[\,\,\,0.01]$ & $0.97^{***}[\,\,\,0.07]$ & $0.96^{***}[\,\,\,0.09]$ & $0.99^{\,\,\,\,\,\,\,\,\,}[\,\,\,0.00]$ & $0.98^{***}[\,\,\,0.02]$ & $0.97^{***}[\,\,\,0.05]$ & $0.96^{***}[\,\,\,0.08]$ & $1.00^{\,\,\,\,\,\,\,\,\,}[\,\,\,0.00]$ \\
friday\_effect & $1.01^{***}[\,\,\,0.03]$ & $1.01^{*\,\,\,\,\,\,}[\,\,\,0.01]$ & $1.01^{***}[\,\,\,0.03]$ & $1.01^{\,\,\,\,\,\,\,\,\,}[\,\,\,0.01]$ & $1.01^{**\,\,\,}[\,\,\,0.01]$ & $1.02^{**\,\,\,}[\,\,\,0.06]$ & $1.01^{\,\,\,\,\,\,\,\,\,}[\,\,\,0.00]$ & $1.01^{***}[\,\,\,0.02]$ & $1.01^{***}[\,\,\,0.02]$ & $1.02^{***}[\,\,\,0.05]$ & $1.01^{*\,\,\,\,\,\,}[\,\,\,0.01]$ & $1.01^{\,\,\,\,\,\,\,\,\,}[\,\,\,0.00]$ \\
contrib\_extra & $0.98^{***}[\,\,\,0.03]$ & $0.99^{*\,\,\,\,\,\,}[\,\,\,0.01]$ & $0.98^{***}[\,\,\,0.05]$ & $1.08^{***}[\,\,\,0.18]$ & $0.99^{.\,\,\,\,\,\,\,}[\,\,\,0.00]$ & $1.00^{\,\,\,\,\,\,\,\,\,}[\,\,\,0.00]$ & $0.98^{***}[\,\,\,0.02]$ & $0.99^{\,\,\,\,\,\,\,\,\,}[\,\,\,0.00]$ & $1.00^{\,\,\,\,\,\,\,\,\,}[\,\,\,0.00]$ & $0.99^{**\,\,\,}[\,\,\,0.02]$ & $0.97^{***}[\,\,\,0.06]$ & $0.95^{***}[\,\,\,0.11]$ \\
src\_churn & $0.99^{**\,\,\,}[\,\,\,0.02]$ & $1.01^{\,\,\,\,\,\,\,\,\,}[\,\,\,0.01]$ & $1.05^{***}[\,\,\,0.15]$ & $0.90^{***}[\,\,\,0.65]$ & $1.00^{\,\,\,\,\,\,\,\,\,}[\,\,\,0.00]$ & $1.00^{\,\,\,\,\,\,\,\,\,}[\,\,\,0.00]$ & $1.05^{***}[\,\,\,0.10]$ & $1.00^{\,\,\,\,\,\,\,\,\,}[\,\,\,0.00]$ & $0.96^{***}[\,\,\,0.10]$ & $0.98^{***}[\,\,\,0.05]$ & $1.01^{\,\,\,\,\,\,\,\,\,}[\,\,\,0.00]$ & $1.00^{\,\,\,\,\,\,\,\,\,}[\,\,\,0.00]$ \\
open\_issue\_num & $0.96^{**\,\,\,}[\,\,\,0.02]$ & $1.15^{***}[\,\,\,0.22]$ & $0.99^{\,\,\,\,\,\,\,\,\,}[\,\,\,0.00]$ & $1.12^{***}[\,\,\,0.13]$ & $1.07^{***}[\,\,\,0.04]$ & $0.90^{***}[\,\,\,0.15]$ & $1.02^{\,\,\,\,\,\,\,\,\,}[\,\,\,0.01]$ & $1.03^{**\,\,\,}[\,\,\,0.01]$ & $0.96^{\,\,\,\,\,\,\,\,\,}[\,\,\,0.01]$ & $0.94^{***}[\,\,\,0.04]$ & $1.07^{**\,\,\,}[\,\,\,0.02]$ & $1.05^{\,\,\,\,\,\,\,\,\,}[\,\,\,0.01]$ \\
sloc & $1.02^{*\,\,\,\,\,\,}[\,\,\,0.01]$ & $1.04^{***}[\,\,\,0.03]$ & $1.04^{***}[\,\,\,0.04]$ & $0.96^{**\,\,\,}[\,\,\,0.04]$ & $0.99^{\,\,\,\,\,\,\,\,\,}[\,\,\,0.00]$ & $1.00^{\,\,\,\,\,\,\,\,\,}[\,\,\,0.00]$ & $1.00^{\,\,\,\,\,\,\,\,\,}[\,\,\,0.00]$ & $0.98^{*\,\,\,\,\,\,}[\,\,\,0.01]$ & $1.07^{***}[\,\,\,0.08]$ & $1.13^{***}[\,\,\,0.33]$ & $0.92^{***}[\,\,\,0.07]$ & $0.94^{***}[\,\,\,0.05]$ \\
files\_deleted & $0.99^{*\,\,\,\,\,\,}[\,\,\,0.01]$ & $0.98^{***}[\,\,\,0.08]$ & $0.96^{***}[\,\,\,0.18]$ & $1.05^{***}[\,\,\,0.28]$ & $0.98^{***}[\,\,\,0.06]$ & $1.00^{\,\,\,\,\,\,\,\,\,}[\,\,\,0.00]$ & $0.97^{***}[\,\,\,0.06]$ & $0.98^{***}[\,\,\,0.04]$ & $0.99^{*\,\,\,\,\,\,}[\,\,\,0.01]$ & $1.02^{***}[\,\,\,0.05]$ & $0.98^{***}[\,\,\,0.04]$ & $0.97^{***}[\,\,\,0.07]$ \\
test\_lines\_per\_kloc & $0.98^{*\,\,\,\,\,\,}[\,\,\,0.01]$ & $0.99^{\,\,\,\,\,\,\,\,\,}[\,\,\,0.00]$ & $1.03^{***}[\,\,\,0.02]$ & $0.93^{***}[\,\,\,0.14]$ & $1.01^{\,\,\,\,\,\,\,\,\,}[\,\,\,0.00]$ & $0.92^{***}[\,\,\,0.17]$ & $0.98^{**\,\,\,}[\,\,\,0.01]$ & $1.02^{*\,\,\,\,\,\,}[\,\,\,0.01]$ & $0.98^{\,\,\,\,\,\,\,\,\,}[\,\,\,0.01]$ & $1.09^{***}[\,\,\,0.22]$ & $0.95^{***}[\,\,\,0.05]$ & $0.94^{***}[\,\,\,0.06]$ \\
followers & $1.00^{\,\,\,\,\,\,\,\,\,}[\,\,\,0.00]$ & $0.96^{***}[\,\,\,0.18]$ & $1.03^{***}[\,\,\,0.06]$ & $1.05^{***}[\,\,\,0.12]$ & $1.04^{***}[\,\,\,0.12]$ & $1.02^{**\,\,\,}[\,\,\,0.04]$ & $1.04^{***}[\,\,\,0.07]$ & $1.01^{\,\,\,\,\,\,\,\,\,}[\,\,\,0.00]$ & $1.07^{***}[\,\,\,0.39]$ & $1.14^{***}[\,\,\,1.40]$ & $1.06^{***}[\,\,\,0.16]$ & $1.04^{***}[\,\,\,0.07]$ \\


has\_comments & \ccell{cellgray}{$0.68^{***}[10.43]$} & \ccell{cellgray}{$0.50^{***}[27.35]$} & - & - & \ccell{cellgray}{$0.65^{***}[10.11]$} & \ccell{cellgray}{$0.52^{***}[24.50]$} & $0.57^{***}[13.21]$ & $0.64^{***}[10.27]$ & $0.66^{***}[11.93]$ & \ccell{cellgray}{$0.63^{***}[14.94]$} & \ccell{cellgray}{$0.62^{***}[\,\,\,9.60]$} & \ccell{cellgray}{$0.55^{***}[13.78]$} \\
other\_comment & $1.24^{***}[\,\,\,5.84]$ & $1.18^{***}[\,\,\,3.60]$ & - & - & $1.23^{***}[\,\,\,4.18]$ & $1.08^{***}[\,\,\,0.71]$ & $1.31^{***}[\,\,\,6.33]$ & $1.23^{***}[\,\,\,4.15]$ & $1.14^{***}[\,\,\,1.92]$ & $1.14^{***}[\,\,\,2.25]$ & $1.22^{***}[\,\,\,3.07]$ & $1.25^{***}[\,\,\,2.84]$ \\
ci\_exists & $1.11^{***}[\,\,\,0.95]$ & $1.19^{***}[\,\,\,2.52]$ & $1.14^{***}[\,\,\,1.64]$ & $1.16^{***}[\,\,\,1.16]$ & - & - & $1.13^{***}[\,\,\,0.76]$ & $1.12^{***}[\,\,\,0.79]$ & $1.09^{***}[\,\,\,0.66]$ & $1.08^{***}[\,\,\,0.64]$ & $1.12^{***}[\,\,\,0.44]$ & $1.17^{***}[\,\,\,0.66]$ \\
num\_comments & $1.12^{***}[\,\,\,0.88]$ & $0.96^{***}[\,\,\,0.11]$ & - & - & $1.01^{.\,\,\,\,\,\,\,}[\,\,\,0.00]$ & $1.18^{***}[\,\,\,1.38]$ & $0.91^{***}[\,\,\,0.37]$ & $1.01^{*\,\,\,\,\,\,}[\,\,\,0.01]$ & $1.13^{***}[\,\,\,0.79]$ & $1.08^{***}[\,\,\,0.31]$ & $1.02^{**\,\,\,}[\,\,\,0.01]$ & $1.07^{***}[\,\,\,0.16]$ \\
comment\_conflict & $1.01^{***}[\,\,\,0.03]$ & $1.01^{***}[\,\,\,0.04]$ & - & - & $1.01^{*\,\,\,\,\,\,}[\,\,\,0.01]$ & $1.02^{**\,\,\,}[\,\,\,0.06]$ & $1.01^{**\,\,\,}[\,\,\,0.02]$ & $1.00^{\,\,\,\,\,\,\,\,\,}[\,\,\,0.00]$ & $1.01^{***}[\,\,\,0.02]$ & $1.01^{***}[\,\,\,0.04]$ & $1.00^{\,\,\,\,\,\,\,\,\,}[\,\,\,0.00]$ & $1.02^{***}[\,\,\,0.04]$ \\
same\_user & - & - & \ccell{cellgray}{$0.56^{***}[29.27]$} & \ccell{cellgray}{$0.42^{***}[41.50]$} & \ccell{cellgray}{$0.51^{***}[32.75]$} & \ccell{cellgray}{$0.59^{***}[23.27]$} & \ccell{cellgray}{$0.49^{***}[24.47]$} & \ccell{cellgray}{$0.49^{***}[36.03]$} & \ccell{cellgray}{$0.55^{***}[32.58]$} & $0.57^{***}[31.20]$ & $0.46^{***}[33.20]$ & $0.46^{***}[28.79]$ \\
inte\_open & - & - & $1.10^{***}[\,\,\,0.61]$ & $1.01^{\,\,\,\,\,\,\,\,\,}[\,\,\,0.01]$ & $1.10^{***}[\,\,\,0.51]$ & $1.04^{***}[\,\,\,0.07]$ & $0.98^{*\,\,\,\,\,\,}[\,\,\,0.01]$ & $0.92^{***}[\,\,\,0.25]$ & $1.18^{***}[\,\,\,2.12]$ & $0.97^{***}[\,\,\,0.05]$ & $1.03^{***}[\,\,\,0.04]$ & $1.25^{***}[\,\,\,2.21]$ \\
inte\_neur & - & - & $1.03^{***}[\,\,\,0.05]$ & $0.99^{\,\,\,\,\,\,\,\,\,}[\,\,\,0.01]$ & $1.06^{***}[\,\,\,0.15]$ & $0.93^{***}[\,\,\,0.24]$ & $0.96^{***}[\,\,\,0.05]$ & $0.93^{***}[\,\,\,0.18]$ & $1.10^{***}[\,\,\,0.58]$ & $0.96^{***}[\,\,\,0.11]$ & $0.98^{*\,\,\,\,\,\,}[\,\,\,0.01]$ & $1.13^{***}[\,\,\,0.57]$ \\
inte\_agree & - & - & $1.00^{\,\,\,\,\,\,\,\,\,}[\,\,\,0.00]$ & $1.05^{***}[\,\,\,0.06]$ & $1.00^{\,\,\,\,\,\,\,\,\,}[\,\,\,0.00]$ & $1.07^{***}[\,\,\,0.14]$ & $1.03^{***}[\,\,\,0.02]$ & $1.01^{\,\,\,\,\,\,\,\,\,}[\,\,\,0.00]$ & $0.98^{***}[\,\,\,0.03]$ & $1.02^{**\,\,\,}[\,\,\,0.02]$ & $1.02^{**\,\,\,}[\,\,\,0.01]$ & $0.92^{***}[\,\,\,0.16]$ \\
inte\_extra & - & - & $1.01^{.\,\,\,\,\,\,\,}[\,\,\,0.00]$ & $0.97^{*\,\,\,\,\,\,}[\,\,\,0.02]$ & $1.02^{***}[\,\,\,0.01]$ & $0.97^{*\,\,\,\,\,\,}[\,\,\,0.03]$ & $1.06^{***}[\,\,\,0.11]$ & $1.07^{***}[\,\,\,0.21]$ & $0.95^{***}[\,\,\,0.14]$ & $1.02^{***}[\,\,\,0.03]$ & $1.04^{***}[\,\,\,0.06]$ & $1.02^{\,\,\,\,\,\,\,\,\,}[\,\,\,0.02]$ \\
inte\_cons & - & - & $1.00^{\,\,\,\,\,\,\,\,\,}[\,\,\,0.00]$ & $1.05^{***}[\,\,\,0.06]$ & $1.01^{\,\,\,\,\,\,\,\,\,}[\,\,\,0.00]$ & $0.98^{.\,\,\,\,\,\,\,}[\,\,\,0.02]$ & $1.01^{\,\,\,\,\,\,\,\,\,}[\,\,\,0.00]$ & $1.00^{\,\,\,\,\,\,\,\,\,}[\,\,\,0.00]$ & $0.99^{**\,\,\,}[\,\,\,0.01]$ & $1.01^{\,\,\,\,\,\,\,\,\,}[\,\,\,0.01]$ & $1.03^{***}[\,\,\,0.02]$ & $0.97^{**\,\,\,}[\,\,\,0.03]$ \\

 \hline \\[-2.2ex]
 Observations & 950,985 & 1,010,937 & 1,152,714 & 809,208 & 1,611,277 & 350,645 & 601,460 & 703,396 & 701,900 & 512,707 & 585,401 & 274,121 \\
 AUC\_train & 0.862 & 0.874 & 0.837 & 0.872 & 0.843 & 0.884 & 0.877 & 0.843 & 0.837 & 0.850 & 0.867 & 0.879 \\[-0.8ex] 
\bottomrule
 \\
\end{tabular} 
\end{table*} 
\end{landscape}
\twocolumn

\end{document}